%% file: aistats/paper.tex
\documentclass[twoside]{article}

\usepackage[preprint]{aistats2027}
\usepackage{aistats/additional}
\usepackage{xcolor}
\input{aistats/math_commands}

\usepackage[round]{natbib}

\input{aistats/macros}

\begin{document}
\etocdepthtag.toc{main}

\twocolumn[

\aistatstitle{Geometric Kernel Diversity: An Axiomatic \\ Diversity Measure in Euclidean Space}

\aistatsauthor{Ditbul Ban \And Younghun Roh \And  Julian Shun \And Vihan Lakshman }

\aistatsaddress{
  MIT CSAIL \\ \texttt{firefly7@mit.edu} \And
  MIT CSAIL \\ \texttt{yhunroh@mit.edu} \And
  MIT CSAIL \\ \texttt{jshun@mit.edu} \And
  MIT CSAIL \\ \texttt{vihan@mit.edu}
}

]

\begin{abstract}
Quantifying the diversity of a set of objects is a fundamental problem in machine learning and statistics. \citet{mironov2025measuring} recently showed that three axioms---monotonicity, uniqueness, and continuity---are crucial for a reliable diversity measure and that none of the standard metrics in the literature satisfy all three.
They also constructed two measures that do satisfy the axioms, but both are NP-hard to compute, leaving open the question of whether efficient axiomatic diversity measures can exist.
We study a Euclidean-restricted version of this question.
We introduce \emph{Geometric Kernel Diversity} (GKD), which associates a radial kernel solid with each point and measures the volume of their union. We show that canonical GKD satisfies all three Mironov--Prokhorenkova axioms on Euclidean-realizable inputs.
For canonical GKD, we give exact polynomial-time algorithms in dimensions $d \le 3$.
In arbitrary dimensions, we give a randomized algorithm that returns a $(1 \pm \varepsilon)$-approximation with probability at least $1-\delta$ in time $O\!\left(\frac{nd}{\varepsilon^2}\log\frac{2}{\delta}\right)$ for $n$ points.
The axiomatic guarantee applies to exact canonical GKD; the randomized algorithm approximates this deterministic quantity.
We further show that the two diversity measures of \citet{mironov2025measuring} remain NP-hard on Euclidean-realizable inputs and, unless $\mathrm{NP} = \mathrm{RP}$, admit no FPRAS even on this restricted domain.
\end{abstract}

\section{Introduction}

Measuring diversity is a fundamental problem in machine learning and statistics, from evaluating generative samples to quantifying dataset coverage and redundancy.
Despite the ubiquity of this task, commonly used diversity measures can behave unintuitively.
\citet{mironov2025measuring} proposed three axioms---monotonicity under nontrivial increases of pairwise distances, uniqueness relative to exact duplicates, and continuity---and showed that none of the common diversity measures in the literature satisfy all three.
They also constructed two measures that do satisfy all three axioms, but both are NP-hard to compute, leaving open the question of whether an axiomatic diversity function can be computationally tractable.

We study a Euclidean-restricted version of this question, motivated in part by settings where objects are represented by learned embeddings.
We introduce \emph{Geometric Kernel Diversity} (GKD): each point induces the under-graph solid of a normalized radial kernel, and GKD is the volume of their union, equivalently the integral of the kernel value determined by the nearest point.
This gives a geometric effective-number notion ranging from complete redundancy toward the support size as the points become widely separated.

For a broad class of kernels, we prove strict monotonicity under pairwise-distance expansions; in particular, our canonical Student-$t_2$ kernel satisfies monotonicity, uniqueness, and continuity on Euclidean-realizable distance matrices.
We then ask whether this axiomatic diversity function can also be computed efficiently. We show that canonical GKD is exactly computable in polynomial time for $d \le 3$ using Voronoi-based algorithms, and in arbitrary dimensions admits a randomized $(1 \pm \varepsilon)$-approximation with probability at least
$1 - \delta$ in $O\!\left(\frac{nd}{\varepsilon^2}\log\frac{2}{\delta}\right)$ time for $n$ points in our randomized real-RAM model.
The axiomatic guarantee applies to exact canonical GKD; the randomized algorithm approximates this deterministic quantity rather than defining a new diversity measure.

In summary, our main contributions are as follows:

\begin{itemize}
\item We introduce Geometric Kernel Diversity (GKD), a new geometric diversity measure in Euclidean space based on the union volume of radial-kernel solids.
For a broad class of radial kernels, we prove that GKD satisfies the monotonicity, uniqueness, and continuity axioms of \citet{mironov2025measuring} on Euclidean-realizable inputs in every dimension.

\item We provide, to our knowledge, the first positive answer to the open problem posed by \citet{mironov2025measuring} in the restricted setting of low-dimensional Euclidean space: for the canonical GKD kernel, we give exact polynomial-time algorithms in dimensions $d \le 3$. The unrestricted Mironov--Prokhorenkova problem remains open.

\item We give a randomized relative-error approximation algorithm for canonical GKD in arbitrary dimensions that returns a $(1 \pm \varepsilon)$-approximation with probability at least $1-\delta$ in time $O\!\left(\frac{nd}{\varepsilon^2}\log\frac{2}{\delta}\right)$ for $n$ points. 
In contrast, we show that the two diversity measures constructed by \citet{mironov2025measuring} remain NP-hard even on Euclidean-realizable inputs and, unless $\mathrm{NP} = \mathrm{RP}$, admit no FPRAS on this restricted domain.
Thus, our results provide the first efficiently computable randomized approximation scheme for an axiomatic diversity measure on Euclidean-realizable inputs. 

\end{itemize}

\section{Related Work}

\paragraph{Axiomatic Diversity.}
\citet{mironov2025measuring} study diversity functions defined on general symmetric nonnegative distance matrices, without requiring the triangle inequality or Euclidean realizability.
They propose three desirable properties. 
\emph{Monotonicity} requires diversity to strictly increase when no pairwise distance decreases and at least one increases. 
\emph{Uniqueness} requires diversity to strictly decrease when a distinct object is replaced by a duplicate of another object.
\emph{Continuity} requires diversity to vary continuously with the pairwise distances. We provide the full formal definitions of these three axioms in Appendix~\ref{app:mironov_axioms}.
\citet{mironov2025measuring} show that none of the measures they consider satisfies all three axioms and construct MultiDimVolume and IntegralMaxClique, which satisfy all three axioms but are NP-hard to compute.
We instead restrict the domain to Euclidean-realizable distance matrices.
This restriction alone does not make their two diversity functions tractable: we show that both remain NP-hard on Euclidean-realizable inputs and that, unless $\mathrm{NP} = \mathrm{RP}$, neither admits an FPRAS on this domain (Appendix~\ref{app:euclidean_hardness}). Moreover, the restriction to the Euclidean domain does not enable existing diversity measures such as the Vendi Score~\citep{friedman2023vendi}, Determinantal Point Processes (DPP)~\citep{wilhelm2018practical}, and Rényi Kernel Entropy (RKE)~\citep{jalali2023information} to satisfy the axioms (Section \ref{sec:exp_axiom_stress}). 

\paragraph{Geometric and volume-based diversity.}
GKD lies within a broader family of geometric upper-envelope constructions.
Its Gaussian specialization coincides, up to normalization and scale, with the Gaussian functionals studied by \citet{mourtada2025} and \citet{cote2025value}, while our definition allows a broader class of radial kernels.
At the other extreme, GKD is the equal-weight, all-maximum specialization of the more general flower-weight framework of \citet{bezdek2008weighted}, which allows more general combinations of radial weight functions.
These works arise in different contexts: in particular, the flower-weight framework is developed in connection with the Kneser-Poulsen problem rather than axiomatic diversity.

Our focus is the Mironov--Prokhorenkova axiomatic question on Euclidean-realizable inputs.
Building on the Gaussian comparison theorem of \citet{mourtada2025}, the Bernstein-Widder representation \citep{widder1931laplace}, and the large-radius strictness theorem of \citet{gorbovickis_strict_2013}, we establish strict monotonicity under arbitrary pairwise-distance expansions for a broader kernel class, a novel result not addressed by prior work. Furthermore, we go beyond prior work in volume-based geometric diversity by designing efficient algorithms. We identify a canonical Student-$t_2$ kernel and give exact algorithms in dimensions $d \le 3$ together with a randomized relative-error approximation in arbitrary dimensions.

\paragraph{Metric-space magnitude.}
\citet{limbeck2024metric} propose magnitude-based measures for evaluating the diversity of latent representations.
For a finite metric space, magnitude is interpreted as an effective number of points across multiple distance scales, and their MagArea measure aggregates this information over scales.
This motivation is closely related to GKD's effective number interpretation, but the theoretical frameworks differ.
They emphasize, among other properties, that adding a genuinely new observation should not decrease diversity and that duplicate observations should be redundant.
They also discuss monotonicity under increases in pairwise distances, but only in a non-strict form and without establishing it in full generality.
In contrast, we study the Mironov--Prokhorenkova axioms, which require strict distance monotonicity and explicitly compare configurations containing duplicates, and establish all three axioms on Euclidean-realizable inputs.

\paragraph{Union-volume approximation.}
\citet{karp1989approx} give the self-adjusting coverage estimator for unions of finite sets, and \citet{bringmann_2010} develop an oracle-based framework for geometric union-volume approximation.
Canonical GKD solids have known unit volume and admit $O(d)$-time exact sampling and membership oracles, enabling the arbitrary-dimensional estimator of Section~\ref{sec:randomized_estimator}; our exact low-dimensional algorithms additionally exploit Voronoi and divergence structure.

\section{Geometric Kernel Diversity} \label{sec:gkd}

Throughout the paper, $\|\cdot\|$ denotes the Euclidean norm, $B_d(\vx, r)$ the Euclidean ball of radius $r$ centered at $\vx$, and $\Vol_k$ the $k$-dimensional Lebesgue measure.
We write $s_{d-1}$ for the surface area of the unit sphere in $\sR^d$.
For a tuple $X=(\vx_1, \dots, \vx_n)$, possibly containing duplicate points, define $\supp(X)$ to be the set of distinct points appearing in $X$. 

\subsection{Definition and Under-Graph Representation}

We begin with a radial kernel that quantifies how strongly a point covers the surrounding space.

\begin{definition}[$d$-admissible radial kernel]
    A function $\phi: [0, \infty) \to [0, \infty)$ is a $d$-admissible radial kernel if it is continuous, strictly decreasing, and
    \[
    0 < \int_0^{\infty} r^{d-1} \phi(r)\,dr < \infty.
    \]
    Its kernel mass is
    \[
    Z_{\phi,d} := \int_{\sR^d} \phi(\|\vz\|)\,d\vz  = s_{d-1} \int_0^{\infty} r^{d-1} \phi(r) \, dr.
    \]
    We call $\phi$ normalized if $Z_{\phi,d}=1$, and assume normalization unless stated otherwise. 
\end{definition}

For a nonempty tuple $X= (\vx_1, \dots, \vx_n) \in (\sR^d)^n$, define
\[
d_X(\vz) := \min_{1 \le i \le n} \|\vz-\vx_i\|.
\]

\begin{definition}[Geometric Kernel Diversity]
    For a normalized $d$-admissible radial kernel $\phi$, define
    \[
    \gD_{\phi,d}(X) := \int_{\sR^d} \phi(d_X(\vz)) \, d\vz.
    \]
\end{definition}

Since $\phi$ is strictly decreasing,
\[
\phi(d_X(\vz)) = \max_{\vx \in \supp(X)} \phi(\|\vz-\vx\|).
\]
For $\vx \in \sR^d$, define its under-graph solid by 
\[
S_{\phi,d}(\vx) := \left\{ (\vz, u) \in \sR^d \times \sR_{\ge 0} \mid 0 \le u \le \phi(\|\vz-\vx\|) \right\}\!.
\]

\begin{proposition}[Under-graph representation and well-definedness] \label{prop:under_graph}
    For every nonempty $X\in(\sR^d)^n$,
    \[
    \gD_{\phi,d}(X) = \Vol_{d+1} \! \left( \bigcup_{\vx \in \supp(X)} S_{\phi,d}(\vx) \right).
    \]
    Moreover, every single-point solid has unit volume: 
    \[
    \Vol_{d+1} \! \left(S_{\phi,d}(\vx)\right) = 1.
    \]
    In particular, $\gD_{\phi,d}(X) < \infty$.
\end{proposition}

\begin{figure}[t]
    \centering
    \includegraphics[width=\linewidth]{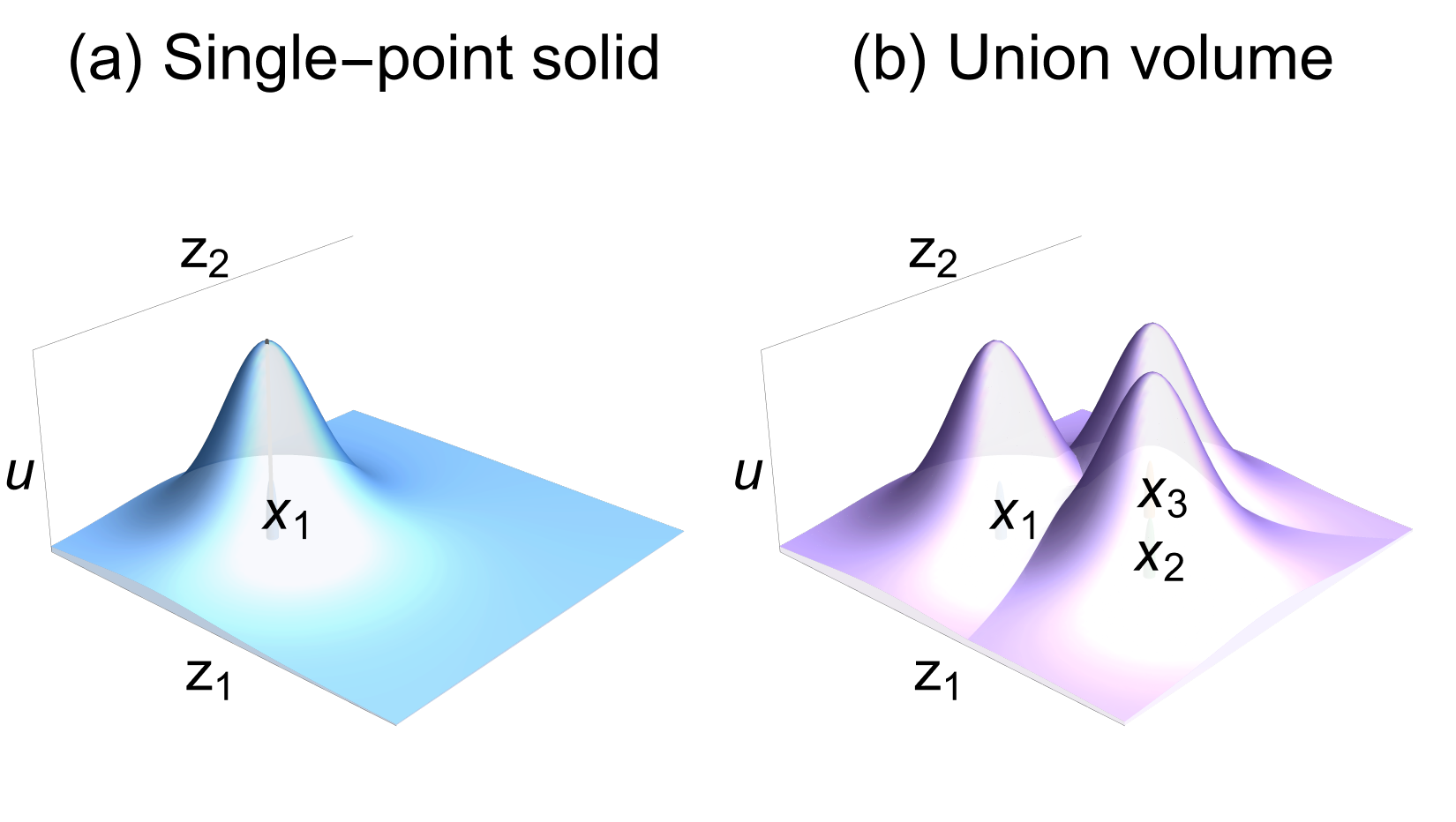}
    \caption{
    \textbf{Union-volume interpretation of GKD in $\mathbb{R}^2$.}
    \textbf{(a)} A point $\vx_i$ induces the under-graph solid $S_{\phi,2}(\vx_i)$.
    \textbf{(b)} For a configuration $X=\{\vx_1,\vx_2,\vx_3\}$, GKD is the volume of the union of the corresponding solids, whose upper boundary is $u=\phi(d_X(\vz))$.
    }
    \label{fig:gkd_union_volume}
\end{figure}

This union-volume viewpoint (Figure~\ref{fig:gkd_union_volume}) will drive both the axiomatic analysis and the computational algorithms below.

\subsection{Basic Properties}

The union-volume representation immediately yields several structural properties of GKD. 

\begin{theorem}[Basic and Coverage Properties] \label{thm:basic_properties}
    Let $d \ge 1$ and let $\phi$ be a normalized $d$-admissible radial kernel. Then, $\gD_{\phi,d}$ is permutation invariant, duplicate invariant, continuous in the input points, and invariant under Euclidean isometries.
    Moreover, adding any point outside the current support strictly increases GKD. With the extension $\gD_{\phi,d}(\varnothing)=0$, its restriction to finite subsets of $\sR^d$ is submodular.
\end{theorem}

These properties are natural from the union-volume viewpoint. Permuting the input points or adding a duplicate leaves the same collection of distinct solids. Adding a genuinely new point introduces a solid with a region of positive volume not covered by the existing ones, so the union volume strictly increases.
Small perturbations of the input points change the corresponding solids continuously, while Euclidean isometries preserve their union volume.
Finally, submodularity expresses the usual diminishing-returns behavior of coverage: as more solids are already present, the additional volume contributed by a new solid can only decrease. The proof is given in Appendix~\ref{app:basic_properties}.

Normalization also gives GKD a natural effective-number interpretation, making its numerical scale directly comparable to the number of distinct points in the configuration.
If $m = |\supp(X)|$, then $\gD_{\phi,d}(X) = 1$ if and only if $m=1$; for $m \ge 2$, 
\[
1 < \gD_{\phi,d}(X) < m,
\]
and the supremum over configurations with $m$ distinct points is $m$, approached as the points become widely separated.
Thus, duplicate points behave like a single effective point, while sufficiently separated points contribute nearly one unit each.
Proposition~\ref{prop:effective_number} in Appendix~\ref{app:effective_number} states and proves these effective-number properties formally.

\subsection{Distance Monotonicity} \label{sec:distance_monotonicity}

We now turn to the main nontrivial property of GKD: increasing pairwise distances should increase diversity.
In the union-volume picture, separating the centers should reduce overlap among their solids and therefore increase the volume of their union.
However, a pairwise-distance expansion is only a comparison between two configurations: in general, there need not be a continuous path of configurations between them along which all pairwise distances are nondecreasing.
Thus, this geometric intuition alone does not give a proof.

For labeled configurations $X = (\vx_1, \dots, \vx_n)$ and $Y = (\vy_1, \dots, \vy_n)$ in $(\sR^d)^n$, write $X \preceq Y$ if
\[
\|\vx_i - \vx_j\| \le \|\vy_i - \vy_j\| \quad \text{for all }i, j,
\]
and write $X \prec Y$ if at least one inequality is strict. We write $X \cong Y$ when all pairwise distances agree, equivalently when the labeled configurations are congruent.

To identify kernels for which monotonicity can be proved, write $\phi(r) = g(r^2)$. A function $g: [0, \infty) \to [0, \infty)$ is \emph{completely monotone} if it is continuous at zero, infinitely differentiable on $(0, \infty)$, and
\[
(-1)^k g^{(k)}(s) \ge 0 \text{ for every $k \ge 0$ and $s > 0$}.
\]
By the Bernstein-Widder theorem \citep{widder1931laplace}, such a function has a unique representation
\[
g(s) = \int_{[0, \infty)} e^{-ts} \, d\mu_g(t)
\]
for a finite nonnegative measure $\mu_g$. Thus, $\phi$ can be expressed as an integral of Gaussian kernels $e^{-tr^2}$ with nonnegative weights.
This is useful because distance monotonicity can first be proved for each Gaussian kernel separately and then integrated over $t$. We say that $g$ has  \emph{accumulating Laplace support} if $\supp(\mu_g)$ has an accumulation point in $(0, \infty)$.

\begin{theorem}[Distance Monotonicity] \label{thm:distance_monotonicity}
    Let $d \ge 1$, and let $\phi$ be a normalized $d$-admissible radial kernel of the form $\phi(r) = g(r^2)$ where $g$ is completely monotone. If $X \preceq Y$, then
    \[
    \gD_{\phi,d}(X) \le \gD_{\phi,d}(Y).
    \]
    If, in addition, $g$ has accumulating Laplace support, then
    \[
    \gD_{\phi,d}(X) = \gD_{\phi,d}(Y) \iff X \cong Y.
    \]
    Consequently, under this additional assumption, $X \prec Y$ implies $\gD_{\phi,d}(X) < \gD_{\phi,d}(Y)$.
\end{theorem}

For a fixed Gaussian scale $t > 0$, define
\[
G_S(t) := \int_{\sR^d} e^{-t d_S(\vz)^2} \, d\vz.
\]
A comparison theorem of \citet{mourtada2025} gives $X \preceq Y \then G_X(t) \le G_Y(t)$ for $t>0$.
Since the representation above gives $\gD_{\phi,d}(S) = \int_{(0, \infty)} G_S(t) \, d\mu_g(t)$, integrating these inequalities proves weak distance monotonicity.

Under the additional accumulating-support assumption, the inequality is strict for every noncongruent expansion; the proof requires a further analytic argument and is given in Appendix~\ref{app:distance_monotonicity}.

\subsection{Mironov--Prokhorenkova Axioms on Euclidean Inputs} \label{sec:axiom_comparison}

The axioms of \citet{mironov2025measuring} are formulated for diversity functions of general symmetric nonnegative distance matrices.
We restrict their framework to distance matrices realizable by points in a prescribed Euclidean space.
Since two labeled Euclidean configurations with the same pairwise distances are congruent, GKD defines a well-defined function of such distance matrices. 

\begin{corollary}[Mironov--Prokhorenkova axioms on Euclidean-realizable distance matrices] \label{cor:mironov_axioms}
    Let $d \ge 1$, and let $\phi$ be a normalized $d$-admissible radial kernel of the form
    \[
        \phi(r) = g(r^2),
    \]
    where $g$ is completely monotone and has accumulating Laplace support.
    For every fixed $n \ge 1$, the diversity function induced by GKD on Euclidean-realizable $n$-point distance matrices is permutation invariant and satisfies the monotonicity, uniqueness, and continuity axioms of \citet{mironov2025measuring}.
\end{corollary}

Each axiom follows from the geometric properties established above. 
Monotonicity is exactly the strict-distance monotonicity result of Theorem~\ref{thm:distance_monotonicity}.
For uniqueness, replacing a genuinely distinct point by a duplicate removes one distinct solid from the union, and Theorem~\ref{thm:basic_properties} shows that this strictly decreases GKD.
Continuity requires one additional step because the axiom is stated in terms of distance matrices, while GKD is defined from point coordinates. If a sequence of Euclidean distance matrices converges, their realizations can be translated and rotated so that the corresponding point coordinates converge. 
Continuity of GKD in the point coordinates then gives continuity as a function of the distance matrix.
The formal construction of the function and proof are given in Appendix~\ref{app:mironov_axioms}.

\section{ The Canonical Kernel and Computation in \texorpdfstring{$d \le 3$}{d <= 3} }

\subsection{The Canonical Kernel} \label{sec:canonical_kernel}

The preceding section establishes the axiomatic properties of GKD for a broad class of radial kernels.
We now specialize to a particular kernel that retains these guarantees while also admitting useful computational structure.

For $d \ge 1$ and $\sigma > 0$, define
\[
\psi_{d,\sigma}(r) := \frac{\Gamma(d/2+1)}{\pi^{d/2}\sigma^d} \left(1 + \frac{r^2}{\sigma^2} \right)^{-(d/2+1)},
\]
where $\Gamma$ denotes the Gamma function.

Then, for $\vz \in \sR^d$, $\vz \longmapsto \psi_{d,\sigma}(\|\vz\|)$ is the density of the spherical $d$-dimensional Student-$t$ distribution with two degrees of freedom and scale matrix $(\sigma^2/2)I_d$, where $I_d$ denotes the $d$-dimensional identity matrix. 
We call $\psi_{d,\sigma}$ the \emph{canonical kernel}, and GKD using this kernel \emph{canonical GKD}.

This choice has several useful properties.
Lemma~\ref{lem:canonical_properties} shows that the canonical kernel satisfies the assumptions of Theorem~\ref{thm:distance_monotonicity}; hence canonical GKD satisfies the Mironov--Prokhorenkova axioms on Euclidean-realizable inputs.
It is also consistent under changes of ambient dimension.
Finally, as we show later, the choice of two degrees of freedom gives especially simple geometric identities that lead to exact low-dimensional algorithms and an efficient sampling procedure for randomized approximation.

\begin{lemma}[Canonical kernel properties] \label{lem:canonical_properties}
    For every $d \ge 1$ and $\sigma > 0$, $\psi_{d,\sigma}$ is a normalized $d$-admissible radial kernel.
    Moreover, $\psi_{d,\sigma}(r) = g_{d,\sigma}(r^2)$, where $g_{d,\sigma}$ is completely monotone and has accumulating Laplace support.
\end{lemma}

Hence Theorem~\ref{thm:distance_monotonicity} and Corollary~\ref{cor:mironov_axioms} apply to canonical GKD.

The parameter $\sigma$ sets the geometric length scale. In particular,
\[
\gD_{\psi_{d,\sigma},d}(X) = \gD_{\psi_{d,1},d}(X/\sigma),
\]
so changing $\sigma$ is equivalent to rescaling the input.
Consequently, the computational arguments below may assume $\sigma=1$ after rescaling. See Lemma~\ref{lem:canonical_scaling} for the formal statement and proof.

\begin{proposition}[Ambient dimension invariance] \label{prop:canonical_ambient_dimension}
    Let $g \ge d \ge 1$, $\sigma > 0$, and let $T: \sR^d \to \sR^g$ be an affine isometric embedding. Then,
    \[
    \gD_{\psi_{g,\sigma},g}(T(X)) = \gD_{\psi_{d,\sigma},d}(X),
    \]
    for every nonempty $X \in (\sR^d)^n$.
\end{proposition}

For two points, canonical GKD has a particularly simple, dimension-independent form. If $r = \|\vx-\vy\|$, then
\[
\gD_{\psi_{d,\sigma},d}(\vx,\vy) = 1 + \frac{r}{\sqrt{r^2 + 4\sigma^2}}.
\]
Thus, the value is $1$ for coincident points and approaches $2$ as the points become widely separated.
This value depends only on the ratio $r/\sigma$, making explicit how $\sigma$ controls the distance scale over which two points are regarded as distinct.
The derivation is given in Proposition~\ref{prop:pairwise_diversity}.

The choice of two degrees of freedom is computationally  useful. 
The canonical density admits a simple transformation to the uniform distribution on the unit ball, which gives the exact sampling procedure used by the randomized algorithm in Section~\ref{sec:randomized_estimator}.
It also admits a simple divergence representation, allowing integrals in the low-dimensional exact algorithms of Section~\ref{sec:exact_algorithms} to be reduced to boundary integrals.
The corresponding identities and their derivations are given in Appendix~\ref{app:canonical_kernel}.

\subsection{Computation Model and Input Conventions} \label{sec:comp_model}

The exact low-dimensional algorithms take as input the point coordinates $X = (\vx_1, \dots, \vx_n) \in (\sR^d)^n$ together with $\sigma>0$, where $d \in \{1, 2, 3\}$ is fixed.
If instead a Euclidean distance matrix is given, coordinates can be reconstructed in polynomial time \citep{alencar2019realizing}; this preprocessing cost is not included in the running times below.

We work in an elementary-function real-RAM model in which real arithmetic, comparisons, square roots, and evaluations of $\arctan$ and $\arccos$ are exact unit-cost operations.
Thus, our running times are real-RAM bounds rather than bit-complexity bounds.
Duplicate points can be removed in $O(n \log n)$ time, and the scaling identity of Lemma~\ref{lem:canonical_scaling} allows the proofs below to assume $\sigma=1$.

\subsection{Exact Algorithms in Low Dimensions} \label{sec:exact_algorithms}

We now give exact algorithms for the canonical GKD in dimensions $d \in \{1, 2, 3\}$.
By Lemma~\ref{lem:canonical_scaling}, the proofs may assume $\sigma=1$.

\begin{theorem}[Exact computation in one dimension] \label{thm:comp_1d}
    Let $x_1, \dots, x_n \in \sR$, $X = (x_1, \dots, x_n)$ and $\sigma > 0$. 
    Then $\gD_{\psi_{1,\sigma},1}(X)$ can be computed in $O(n \log n)$ time.
    If $x_1 \le x_2 \le \dots \le x_n$ and $\Delta_i = x_{i+1} - x_i$, then
    \[
    \gD_{\psi_{1,\sigma},1}(X) = 1 + \sum_{i=1}^{n-1} \frac{\Delta_i}{\sqrt{\Delta_i^2 + 4\sigma^2}}.
    \]
\end{theorem}

After sorting the points, decomposing the line according to the nearest input point and integrating over the resulting intervals yields the formula above.
The formula then can be evaluated in $O(n)$ time, so the overall running time is $O(n \log n)$.
See Appendix~\ref{app:comp_1d}.

The same nearest-point decomposition is given in higher dimensions by the Euclidean Voronoi diagram. 
Inside each Voronoi cell, the nearest input point is fixed, so GKD decomposes into one integral per cell. 
For the canonical kernel, the divergence theorem converts each cell integral into integrals over its boundary faces, reducing the dimension of the remaining integration by one.
The full reduction is given in Appendix~\ref{app:voronoi_divergence}.

\begin{theorem}[Exact computation in two dimensions] \label{thm:comp_2d}
    For $X \in (\sR^2)^n$ and $\sigma > 0$, $\gD_{\psi_{2,\sigma},2}(X)$ can be computed in $O(n \log n)$ time in the elementary real-RAM model.
\end{theorem}

In two dimensions, Voronoi boundaries are one-dimensional, and each resulting boundary integral has an elementary closed form.
A planar Voronoi diagram can be constructed in $O(n \log n)$ time \citep{fortune_sweepline_1987}, after which all boundary contributions can be evaluated in $O(n)$ total time.
The construction is given in Appendix~\ref{app:comp_2d}.

\begin{theorem}[Exact computation in three dimensions] \label{thm:comp_3d}
    For $X \in (\sR^3)^n$ and $\sigma > 0$, $\gD_{\psi_{3,\sigma},3}(X)$ can be computed in $O(n^2)$ time in the elementary real-RAM model.
\end{theorem}

In three dimensions, the boundary terms are integrals over polygonal Voronoi faces. Appendix~\ref{app:planar_face_integration} gives an elementary boundary formula for evaluating each such face integral.
A three-dimensional Voronoi diagram can have $O(n^2)$ total complexity and can be constructed in $O(n^2)$ time \citep{chazelle_optimal_1993}; Appendix~\ref{app:comp_3d} shows that the total work needed to evaluate all face contributions is also $O(n^2)$.

\paragraph{Higher dimensions.}
For $d \ge 4$, this approach becomes less attractive for two reasons.
First, the total size of a $d$-dimensional Voronoi diagram---the number of faces of all dimensions that may need to be processed---can be $\Theta(n^{\lceil d/2 \rceil})$ in the worst case.
For fixed $d$, this is still polynomial in $n$, but the exponent grows with $d$; without a fixed-dimensional restriction, an $n$-point Euclidean distance matrix may require dimension as large as $n-1$.
Second, the divergence reduction leaves $(d-1)$-dimensional integrals over Voronoi faces, and we do not have comparably simple exact evaluations for these higher-dimensional integrals. 
We therefore turn to randomized approximation, which avoids explicit Voronoi construction and applies in arbitrary dimensions.

\section{Estimator in Arbitrary Dimensions} \label{sec:randomized_estimator}

Canonical GKD itself is the deterministic diversity function defined in the previous sections, and its axiomatic guarantees concern this exact function.
In arbitrary dimensions, rather than compute its value exactly, we use a randomized algorithm to approximate it. 
The randomized estimator is a computational procedure for approximating this deterministic function and is not itself an axiomatic diversity measure; its output may vary across runs and does not satisfy the continuity axioms.
Our estimator is, to our knowledge, the first efficient randomized relative-error approximation scheme for a diversity function that satisfies all three Mironov--Prokhorenkova axioms since the prior two diversity measures constructed by \citet{mironov2025measuring}, namely IntegralMaxClique and MultiDimVolume, do not admit an FPRAS, unless NP=RP (Appendix~\ref{sec:hardness-randomized-apx}). 

We augment the elementary real-RAM model of Section~\ref{sec:comp_model} with exact unit-cost independent sampling from $\Unif(0, 1)$, $\gN(0, 1)$, and finite uniform distributions.
The sampling and membership oracles needed below are constructed in Appendix~\ref{app:estimator}.

For $X = (\vx_1, \dots, \vx_n) \in (\sR^d)^n$, let
\[
S_i := S_{\psi_{d,\sigma},d}(\vx_i) \subseteq \sR^{d+1}.
\]
By Proposition~\ref{prop:under_graph},
\[
\gD_{\psi_{d,\sigma},d}(X) = \Vol_{d+1} \! \left( \bigcup_{i=1}^n S_i \right), \quad \Vol_{d+1}(S_i) = 1.
\]
Thus, estimating canonical GKD is an instance of estimating the volume of a union of $n$ equal-volume sets.
We specialize the self-adjusting coverage estimator of \citet{karp1989approx} to this setting.

A trial first chooses $i$ uniformly from $\{1, \dots, n\}$ and samples a point $Y$ uniformly from $S_i$.
It then repeatedly chooses independent uniform indices $j$ and tests whether $Y \in S_j$, stopping at the first success.

For a point $y$ in the union, let
\[
c(\vy) := \left|\left\{j \in \{1, \dots, n\} \mid \vy \in S_j \right\}\right|
\]
be the number of solids covering $y$.
Points with larger $c(y)$ are more likely to be sampled: the density of $Y$ at $y$ is $c(y)/n$.
On the other hand, once $Y = y$ has been sampled, a random membership query succeeds with probability $c(y)/n$.
Hence, if $L$ is the number of membership queries made in one trial,
\[
L \mid (Y = \vy) \sim \Geom \! \left( \frac{c(\vy)}{n} \right),
\]
and therefore
\[
\E[L \mid Y = y] = \frac{n}{c(y)}.
\]
These two effects cancel: regions covered by many solids are sampled more often, but require few membership queries on average.
Consequently,
\[
\begin{aligned}
\E[L] &= \int_{\cup_i S_i} \frac{n}{c(\vy)} \frac{c(\vy)}{n} \, d\vy \\
&= \Vol_{d+1}\!\left(\bigcup_{i=1}^n S_i\right) = \gD_{\psi_{d,\sigma},d}(X).
\end{aligned}
\]
Thus, the expected length of one trial is exactly the canonical GKD value.
The fixed-budget estimator below turns this observation into a relative-error approximation with high probability.

To obtain a high-probability relative-error guarantee, we use a fixed total number of membership queries. 
For $\eps, \delta \in (0, 1)$, set 
\[
B := \left\lceil \frac{8(1+\eps)n \log(2/\delta)}{\eps^2} \right\rceil.
\]
Run consecutive trials until the $B$-th membership query, and let $N_B$ be the number of trials completed within this budget. 
Since the expected length of each trial is the GKD value, $N_B$ is approximately $B/\gD_{\psi_{d,\sigma},d}(X)$, motivating the estimator
\[
\widehat D = \begin{cases}
    \dfrac{B}{N_B} & (N_B > 0), \\
    n & (N_B = 0).
\end{cases}
\]
The case $N_B = 0$ occurs exactly when the first trial has not finished within the $B$-query budget.
We then set $\widehat D = n$ only to make the estimator well-defined.
With probability at least $1-\delta$, the analysis guarantees that $N_B > 0$ and the required relative-error bound, so this fallback value does not affect the theorem.

\begin{theorem}[Randomized relative approximation] \label{thm:estimator}
    For every nonempty $X \in (\sR^d)^n$, $d \ge 1$, and $\eps, \delta \in (0, 1)$,
    \[
    \Pr\! \left[
    \left| \widehat{D} - \gD_{\psi_{d,\sigma}, d}(X) \right| \le \eps \, \gD_{\psi_{d,\sigma},d}(X) \right]
    \ge 1 - \delta.
    \]
    In the randomized elementary real-RAM model, the running time is
    \[
    O \! \left( \frac{nd}{\eps^2} \log \frac{2}{\delta} \right).
    \]
\end{theorem}
The fixed-budget analysis of the self-adjusting coverage estimator gives the stated relative-error guarantee.
Sampling uniformly from a canonical solid and testing membership both take $O(d)$ time, so the $B$ membership queries and at most $B$ trial initializations require $O(Bd)$ total time.
The full analysis and oracle constructions are given in Appendix~\ref{app:estimator}.

\paragraph{Comparison with prior diversity measures.}
The Euclidean restriction does not make the two diversity measures of \citet{mironov2025measuring} tractable.
In Appendix~\ref{app:euclidean_hardness}, we show that MultiDimVolume and IntegralMaxClique remain NP-hard even on Euclidean-realizable inputs and, unless $\mathrm{NP} = \mathrm{RP}$, neither admits an FPRAS on this restricted domain.
The reductions encode arbitrary graphs as small perturbations of a regular simplex.

\section{Axiom Stress Tests}

\citet{mironov2025measuring} show that standard diversity measures violate their axioms on general symmetric distance matrices, a domain that includes matrices violating the triangle inequality and matrices not realizable by any point configuration. A priori, these axiomatic failures could be artifacts of that broader domain, and restricting to Euclidean inputs might be enough to restore the axioms. This section shows that this is not the case. The violations persist both on simple, explicitly constructed low-dimensional Euclidean configurations and under Euclidean-realizable perturbations of real learned representations (deferred to Appendix~\ref{app:exp_imagenet}). These experiments indicate that the Euclidean restriction does not by itself resolve the tension between the axioms and efficient computation.

\subsection{Comparison Measures}
\label{sec:exp_setup}

We consider several commonly used diversity measures in addition to GKD: average pairwise distance, the Vendi Score \citep{friedman2023vendi}, determinant-based DPP diversity \citep{mironov2025measuring,wilhelm2018practical}, and Rényi kernel entropy (RKE) \citep{jalali2023information}.  Full definitions are given in Appendix~\ref{app:exp_measures}. Our goal in this section is not to argue that one diversity measure is superior to another for a particular application, but rather to characterize their behavior in relation to the Mironov--Prokhorenkova axioms.

For the Vendi Score, DPP, and RKE, we use the RBF similarity matrix
\[
K_{ij} := \exp \! \left( -\frac{\|x_i-x_j\|^2}{2h^2} \right),
\]
with a common bandwidth $h$ within each experiment.

All of these comparison measures are continuous on Euclidean inputs, so our experiments focus on the monotonicity and uniqueness axioms that they can violate.

\subsection{Controlled Euclidean Stress Tests}
\label{sec:exp_axiom_stress}

We first use small Euclidean configurations to show that several standard comparison measures can violate the monotonicity and uniqueness axioms.
These examples are designed to isolate a single axiom rather than to model a particular application.
We set $\sigma = h = 1$ and compute canonical GKD exactly.

\paragraph{Distance monotonicity.}
Consider three points with
\[
d_{12} = d_{13} = 1, \quad d_{23} = c \in [1.76,1.99].
\]
Increasing $c$ changes exactly one pairwise distance while leaving the other two fixed, so strict distance monotonicity requires the diversity score to increase.
Canonical GKD does so, while the Vendi score and determinant-based DPP diversity both decrease along this family.

\paragraph{Uniqueness.}
Consider the one-dimensional configuration
\[
X(t) = \left(0,0,\frac12,\frac12-\frac{t}{4}\right), \quad t\in[0,1].
\]
At $t=0$, the third and fourth points coincide.
For every $t > 0$, the fourth point becomes distinct while the duplicate pair at the origin remains unchanged.
The uniqueness axiom therefore requires $D(X(t))>D(X(0))$ for every $t>0$. 
Canonical GKD increases as required.
Average pairwise distance, the Vendi Score, and RKE decrease on this family, while the determinant-based DPP score remains zero.

These examples show that axiom violations can occur even in simple low-dimensional Euclidean configurations.
Appendix~\ref{app:exp_controlled} gives the explicit point configurations, plots of the diversity scores along both perturbations, and numerical endpoint values for all measures.

Additional stress tests under Euclidean-realizable perturbations around ImageNet/DINOv2 embeddings are given in Appendix~\ref{app:exp_imagenet}.

\section{Conclusion} \label{sec:discussion}

We studied the question of \citet{mironov2025measuring} under the restriction that the input distance matrix is Euclidean-realizable, and introduced Geometric Kernel Diversity (GKD).
For a broad class of radial kernels, GKD satisfies monotonicity, uniqueness, and continuity on this domain.
For the canonical Student-$t_2$ kernel, GKD is exactly computable in $O(n \log n)$ time in one and two dimensions  and in $O(n^2)$ time in three dimensions, while admitting a randomized $(1 \pm \eps)$-approximation in arbitrary dimensions.
We also show that the two diversity measures of \citet{mironov2025measuring} remain NP-hard on Euclidean-realizable inputs and, unless $\mathrm{NP} = \mathrm{RP}$, do not admit an FPRAS even under this restriction.

Several questions remain open.
Most importantly, it is still unknown whether a diversity function satisfying all three axioms can be computed efficiently on unrestricted symmetric distance matrices. 
For GKD itself, it remains open whether canonical GKD admits efficient exact algorithms in dimensions $d \ge 4$, and whether other kernels retaining the axiomatic guarantees lead to simpler or faster computation. Each of these questions represents exciting directions for future work, building on the results we present in this paper. 

\subsection*{AI use statement}

In this work, we used generative AI tools for generating synthetic data sets; helping develop theoretical models or conceptual frameworks; summarizing or analyzing existing literature; formulating mathematical claims; providing critical ingredients for proving mathematical claims; assisting in the writing of proofs; drafting parts of the paper; improving the writing of the paper; proposing or refining hypotheses; designing or providing feedback on research methodology or experiments; implementing methods; cleaning and reformatting datasets; creating or modifying figures or images; suggesting experimental parameters; creating or editing software code; creating artifacts; and interpreting results. 
We have not used generative AI tools for any other applicable tasks with required disclosure. Assisting with translation and supporting qualitative and thematic data analysis were not applicable to this work.

We reviewed all AI-assisted work. Mathematical claims and proofs developed with AI assistance were checked by the authors. AI-assisted experimental and data-processing code was tested through exact calculations, reproducibility checks, and comparisons with recorded experimental outputs, as applicable. Suggested literature and citations were checked against the corresponding source material. We take responsibility for the final content of this work, including text, claims, or artifacts produced with the aid of generative AI.

\bibliography{aistats/references}

\clearpage
\appendix
\etocdepthtag.toc{appendix}
\thispagestyle{empty}

\onecolumn
\aistatstitle{Supplementary Materials}

\etocsettagdepth{main}{none}
\etocsettagdepth{appendix}{subsection}
\etocsettocstyle{\section*{Contents}}{}
\tableofcontents

\newpage
\section{Properties of Geometric Kernel Diversity} \label{app:gkd_properties}

\subsection{Under-Graph Representation}

\begin{proof}[Proof of Proposition~\ref{prop:under_graph}]

For every $\vz \in \sR^d$, since $\phi$ is strictly decreasing,
\[
\phi(d_X(\vz)) = \max_{\vx \in \supp(X)} \phi(\|\vz-\vx\|).
\]
Hence
\[
\begin{aligned}
    \left\{(\vz,u) \in \sR^d \times \sR_{\ge 0} \mid 0 \le u \le \phi(d_X(\vz)) \right\} &= \bigcup_{\vx \in \supp(X)} \left\{ (\vz, u) \in \sR^d \times \sR_{\ge 0} \mid 0 \le u \le \phi(\|\vz-\vx\|)\right\} \\
    &= \bigcup_{\vx \in \supp(X)} S_{\phi,d}(\vx).
\end{aligned}
\]
These sets are measurable because $d_X$ and $\phi$ are continuous. Therefore, by Tonelli's theorem,
\[
\begin{aligned}
    \Vol_{d+1} \! \left( \bigcup_{\vx \in \supp(X)} S_{\phi,d}(\vx) \right) &= \int_{\sR^d} \int_0^{\infty} \1_{u \le \phi(d_X(\vz))} \, du \, d\vz \\
    &= \int_{\sR^d} \phi(d_X(\vz)) \, d\vz \\
    &= \gD_{\phi,d}(X).
\end{aligned}
\]
For every $\vx \in \sR^d$, translation invariance of Lebesgue measure and normalization of $\phi$ give
\[
\Vol_{d+1}(S_{\phi,d}(\vx)) = \int_{\sR^d} \phi(\|\vz-\vx\|) \, d\vz = Z_{\phi,d} = 1.
\]
Since $X$ is nonempty, the union contains at least one such unit-volume solid, and hence
\[
\gD_{\phi,d}(X) \ge 1.
\]
On the other hand, subadditivity of Lebesgue measure gives
\[
\begin{aligned}
    \gD_{\phi,d}(X) &= \Vol_{d+1} \! \left( \bigcup_{\vx \in \supp(X)} S_{\phi,d}(\vx) \right) \le \sum_{\vx \in \supp(X)} \Vol_{d+1} \left( S_{\phi,d}(\vx) \right) = |\supp(X)|.
\end{aligned}
\]
Finally, $|\supp(X)| \le n$, and therefore,
\[
1 \le \gD_{\phi,d}(X) \le |\supp(X)| \le n.
\]
In particular, $\gD_{\phi,d}(X)$ is finite.
\end{proof}

\subsection{Basic and Coverage Properties}
\label{app:basic_properties}

\begin{proof}[Proof of Theorem~\ref{thm:basic_properties}]
\leavevmode

\paragraph{Permutation and duplicate invariance.}
The nearest-distance function depends only on the set of distinct input points:
\[
d_X(\vz) = \min_{\vx \in \supp(X)} \|\vz-\vx\|.
\]
Permuting the entries of $X$, or adding another copy of an existing point, therefore leaves $d_X$ unchanged and hence leaves $\gD_{\phi,d}(X)$ unchanged.

\medskip
\paragraph{Strict addition monotonicity.} 
Let $X$ be nonempty and let $\vw \notin \supp(X)$. Set
\[
\ell := d_X(\vw) > 0.
\]
For every $\vz \in B_d(\vw, \ell/3)$ and every $\vx \in \supp(X)$,
\[
\|\vz-\vx\| \ge \|\vw - \vx\| - \|\vz - \vw\| \ge \frac{2\ell}{3},
\]
whereas
\[
\|\vz-\vw\| \le \frac{\ell}{3}.
\]
Hence,
\[
d_{X \cup \{\vw\}}(\vz) = \|\vz-\vw\| < d_X(\vz)
\]
throughout $B_d(\vw, \ell/3)$. Since $\phi$ is strictly decreasing,
\[
\phi(d_{X \cup \{\vw\}}(\vz)) > \phi(d_X(\vz))
\]
on this ball. Moreover, adding a point can only decrease the nearest-point distance, so for every $\vz \in \sR^d$,
\[
d_{X \cup \{\vw\}}(\vz) \le d_X(\vz)
\]
and therefore
\[
\phi(d_{X \cup \{\vw\}}(\vz)) \ge \phi(d_X(\vz)).
\]
Since $B_d(\vw, \ell/3)$ has positive Lebesgue measure, integrating these inequalities gives
\[
\gD_{\phi,d}(X \cup \{\vw\}) > \gD_{\phi,d}(X).
\]
For $X = \varnothing$, the extension $\gD_{\phi,d}(\varnothing) = 0$ gives
\[
\gD_{\phi,d}(\{\vw\}) = 1 > 0.
\]

\medskip
\paragraph{Continuity.} 
Suppose 
\[
X^{(k)} = (\vx_1^{(k)}, \vx_2^{(k)}, \dots, \vx_n^{(k)}) \longrightarrow X = (\vx_1, \dots, \vx_n)
\]
in $(\sR^d)^n$. For every fixed $\vz \in \sR^d$, continuity of the minimum gives
\[
d_{X^{(k)}}(\vz) \longrightarrow d_X(\vz),
\]
and therefore
\[
\phi(d_{X^{(k)}} (\vz)) \longrightarrow \phi(d_X(\vz)).
\]
Since $X^{(k)} \to X$, there exists $R>0$ such that $\|\vx_i^{(k)}\| \le R$ for all $i$ and $k$. Therefore,
\[
d_{X^{(k)}}(\vz) \ge \max\{0, \|\vz\|-R\},
\]
so, because $\phi$ is decreasing,
\[
0 \le \phi(d_{X^{(k)}}(\vz)) \le h(\vz) :=\phi(\max\{0, \|\vz\|-R\}).
\]
The function $h$ is integrable. Indeed, it is bounded by $\phi(0)$ on $B_d(\vzero, 2R)$, while for $\|\vz\| > 2R$,
\[
h(\vz) \le \phi(\|\vz\|/2),
\]
and
\[
\int_{\sR^d} \phi(\|\vz\|/2) \, d\vz = 2^d \int_{\sR^d} \phi(\|\vu\|) \, d\vu = 2^d.
\]
The dominated convergence theorem therefore yields
\[
\gD_{\phi,d}(X^{(k)}) \longrightarrow \gD_{\phi,d}(X).
\]

\medskip
\paragraph{Euclidean Isometry Invariance.} Let $T: \sR^d \to \sR^d$ be a Euclidean isometry. Since $T$ preserves distances,
\[
d_{T(X)} (T\vz) = d_X(\vz).
\]
It also preserves Lebesgue measure, so the change of variables $\vw = T\vz$ gives
\[
\begin{aligned}
    \gD_{\phi,d}(T(X)) &= \int_{\sR^d} \phi(d_{T(X)}(\vw)) \, d\vw \\
    &= \int_{\sR^d} \phi(d_{T(X)}(T\vz)) \, d\vz\\
    &= \int_{\sR^d} \phi(d_X(\vz)) \, d\vz = \gD_{\phi,d}(X).
\end{aligned}
\]

\medskip
\paragraph{Submodularity.} For a finite set $C \subseteq \sR^d$, define
\[
U_C := \bigcup_{\vx \in C} S_{\phi,d}(\vx), \quad U_\varnothing := \varnothing.
\]
By Proposition~\ref{prop:under_graph},
\[
\gD_{\phi,d}(C) = \Vol_{d+1}(U_C).
\]
Let $A \subseteq B$ be finite sets and let $\vw \notin B$.
Then $U_A \subseteq U_B$, and
\[
\begin{aligned}
    \gD_{\phi,d}(A \cup \{\vw\}) - \gD_{\phi,d}(A) &=
    \Vol_{d+1} \! \left(S_{\phi,d}(\vw) \setminus U_A \right), \\
    \gD_{\phi,d}(B \cup \{\vw\}) - \gD_{\phi,d}(B) &=
    \Vol_{d+1} \! \left(S_{\phi,d}(\vw) \setminus U_B \right).
\end{aligned}
\]
Since
\[
S_{\phi,d}(\vw) \setminus U_A \supseteq S_{\phi,d}(\vw) \setminus U_B,
\]
we obtain
\[
\gD_{\phi,d}(A \cup \{\vw\}) - \gD_{\phi,d}(A) \ge 
\gD_{\phi,d}(B \cup \{\vw\}) - \gD_{\phi,d}(B).
\]
Thus, the restriction of $\gD_{\phi,d}$ to finite sets is submodular.
\end{proof}

\subsection{Effective-Number Interpretation} \label{app:effective_number}

The normalization of each single-point solid gives GKD a natural effective-number interpretation.

\begin{proposition}[Effective-number interpretation]
\label{prop:effective_number}
Let $d \ge 1$, let $\phi$ be a normalized $d$-admissible radial kernel, and let $X=(\vx_1,\ldots,\vx_n)\in(\sR^d)^n$ be nonempty.
Define
\[
    m:=|\supp(X)|.
\]
Then
\[
    \gD_{\phi,d}(X)=1 \iff m=1.
\]
If $m\ge2$, then
\[
    1 < \gD_{\phi,d}(X) < m.
\]
Moreover, for every integer $m\ge1$,
\[
    \sup\left\{ \gD_{\phi,d}(X) \;\middle|\; X \subset \sR^d,\ |X|=m
    \right\} = m.
\]
For $m \ge 2$, this supremum is not attained.
\end{proposition}
\begin{proof}

If $m = 1$, then duplicate invariance and normalization give
\[
\gD_{\phi,d}(X) = 1.
\]
Conversely, if $m \ge 2$, start from any one point of $\supp(X)$ and add the remaining distinct points one at a time.
Every addition is strict by Theorem~\ref{thm:basic_properties}, so
\[
\gD_{\phi,d}(X) > 1.
\]
It remains to prove the strict upper bound. By Proposition~\ref{prop:under_graph},
\[
\gD_{\phi,d}(X) = \Vol_{d+1} \! \left( \bigcup_{\vx \in \supp(X)} S_{\phi,d}(\vx) \right) \le m.
\]
Suppose $m \ge 2$, and choose distinct $\vx, \vy \in \supp(X)$.
Let
\[
\vc := \frac{\vx+\vy}{2}, \quad a := \phi \! \left( \frac{\|\vx-\vy\|}{2} \right).
\]
Since $\phi$ is strictly decreasing and nonnegative, $a>0$.
By continuity, there exists an open neighborhood $U$ of $\vc$ such that 
\[
\phi(\|\vz-\vx\|) > \frac a2, \quad \phi(\|\vz-\vy\|) > \frac a2
\]
for every $\vz \in U$.
Hence
\[
U \times [0, a/2] \subseteq S_{\phi,d}(\vx) \cap S_{\phi,d}(\vy),
\]
so two of the union-volume solids overlap on a set of positive $(d+1)$-dimensional measure.
Therefore,
\[
\gD_{\phi,d}(X) < m.
\]
Finally, fix $m \ge 1$ and $\eps > 0$. Since $\phi$ is normalized, choose $R > 0$ such that
\[
\int_{B_d(\vzero, R)} \phi(\|\vz\|) \, d\vz > 1 - \frac{\eps}{m}.
\]
Let $\ve_1$ be the first standard basis vector and set
\[
\vx_i := 3Ri \, \ve_1, \quad i=1, \dots, m.
\]
The balls $B_d(\vx_i, R)$ are pairwise disjoint, and every point of $B_d(\vx_i, R)$ is closer to $\vx_i$ than to any other $\vx_j$.
Therefore,
\[
\begin{aligned}
    \gD_{\phi,d}(\vx_1, \dots, \vx_m) & \ge \sum_{i=1}^m \int_{B_d(\vx_i, R)} \phi(\|\vz-\vx_i\|) \, d\vz \\
    &= m \int_{B_d(\vzero, R)} \phi(\|\vz\|) \, d\vz \\
    &> m - \eps.
\end{aligned}
\]
Together with the universal upper bound $\gD_{\phi,d} \le m$, this shows
\[
\sup \left \{ \gD_{\phi,d}(X) \;\middle|\; X \subset \sR^d, \, |X| = m \right\} = m.
\]
When $m \ge 2$, the strict upper bound above shows that the supremum is not attained.

\end{proof}

\subsection{Distance Monotonicity}
\label{app:distance_monotonicity}

For a nonempty finite configuration $S \subset \sR^d$ and $t>0$, define its Gaussian functional by
\[
G_S(t) := \int_{\sR^d} e^{-t d_S(\vz)^2} d\vz.
\]
We first record the Gaussian comparison supplied by the nonconvex Wills functional.

\begin{lemma}[Gaussian Comparison] \label{lem:gaussian_comparison}
    Let $X, Y \in (\sR^d)^n$ satisfy $X \preceq Y$. Then, for every $t > 0$, 
    \[
    G_X(t) \le G_Y(t).
    \]
\end{lemma}
\begin{proof}
    Consider
    \[
    \tau: \supp(Y) \to \supp(X), \quad \tau(\vy_i) := \vx_i.
    \]
    This map is well-defined: if $\vy_i = \vy_j$, then
    \[
    \|\vx_i - \vx_j\| \le \|\vy_i - \vy_j\| = 0,
    \]
    so $\vx_i = \vx_j$. It is surjective because every point of $\supp(X)$ is $\vx_i = \tau(\vy_i)$ for some $i$. Moreover,
    \[
    \|\tau(\vy_i) - \tau(\vy_j)\| = \|\vx_i - \vx_j\| \le \|\vy_i - \vy_j\|.
    \]
    Thus, $\tau$ is distance-contracting.
    
    By the comparison theorem for the nonconvex Wills functional \citep[Theorem~4.1, Eq.~(2.16)]{mourtada2025},
    \[
    \int_{\sR^d} e^{-\pi d_X(\vz)^2} d\vz \le \int_{\sR^d} e^{-\pi d_Y(\vz)^2} d\vz.
    \]
    To obtain the result for arbitrary $t > 0$, define 
    \[
    a := \sqrt{\frac{t}{\pi}}.
    \]
    Since $X \preceq Y$, we also have $aX \preceq aY$. Applying the preceding inequality to $aX$ and $aY$, and using the substitution $\vw = a\vz$, gives
    \[
    \begin{aligned}
        \int_{\sR^d} e^{-\pi d_{aX}(\vw)^2} d\vw &= a^d \int_{\sR^d} e^{-\pi a^2 d_X(\vz)^2} d\vz \\
        &= a^d G_X(t),
    \end{aligned}
    \]
    and similarly
    \[
    \int_{\sR^d} e^{-\pi d_{aY}(\vw)^2} d\vw = a^d G_Y(t).
    \]
    Dividing by $a^d$ yields
    \[
    G_X(t) \le G_Y(t).
    \]
\end{proof}

\begin{lemma}[Analyticity of the Gaussian functional] 
\label{lem:gaussian_functional_analytic}
    For every nonempty finite configuration $S \subset \sR^d$, the function $G_S$ is real analytic on $(0, \infty)$.
\end{lemma}
\begin{proof}
    For complex $w$ with $\Re(w) > 0$, define
    \[
    G_S(w) := \int_{\sR^d} e^{-w d_S(\vz)^2} d\vz.
    \]
    Let $K$ be a compact subset of the half-plane $\{w \in \sC \mid \Re(w) > 0\}$. Choose $\eta > 0$ such that $\Re(w) \ge \eta$ for every $w \in K$. Choose $R > 0$ so that $S \subset B_d(\vzero, R)$. For $\|\vz\| \ge 2R$,
    \[
    d_S(\vz) \ge \|\vz\|-R \ge \frac{\|\vz\|}{2}.
    \]
    Therefore,
    \[
    \left| e^{-w d_S(\vz)^2} \right| \le e^{-\eta d_S(\vz)^2} \le e^{-\eta \|\vz\|^2/4}
    \]
    outside $B_d(\vzero, 2R)$, while the integrand is bounded by $1$ inside this ball. Thus, the integrand is dominated, uniformly over $w \in K$, by an integrable function.

    For every fixed $\vz$, the integrand is holomorphic in $w$, and the preceding bounds give an integrable dominating function locally uniformly in $w$. It follows that $G_S$ is holomorphic on the half-plane $\Re(w) > 0$. Therefore, its restriction to $(0, \infty)$ is real analytic.
\end{proof}

\begin{proof}[Proof of Theorem~\ref{thm:distance_monotonicity}]
    \leavevmode
    
    \paragraph{Gaussian-mixture representation and weak monotonicity.}
    By the Bernstein-Widder theorem, there exists a finite nonnegative Borel measure $\mu_g$ on $[0, \infty)$ such that
    \[
    g(s) = \int_{[0, \infty)} e^{-ts} d\mu_g(t).
    \]
    We first observe that
    \[
    \mu_g(\{0\}) = 0.
    \]
    Indeed, if $\mu_g(\{0\}) > 0$, then,
    \[
    g(s) \ge \mu_g(\{0\}) > 0
    \]
    for every $s \ge 0$. Hence, for every nonempty configuration $S$,
    \[
    \gD_{\phi,d}(S) = \int_{\sR^d} g(d_S(\vz)^2) d\vz = \infty,
    \]
    contradicting the assumption that $\phi$ is a $d$-admissible kernel. Therefore, 
    \[
    g(s) = \int_{(0, \infty)} e^{-ts} d\mu_g(t).
    \]
    Using Tonelli's theorem,
    \[
    \begin{aligned}
        \gD_{\phi,d}(S) &= \int_{\sR^d} g(d_S(\vz)^2) d\vz \\
        &= \int_{\sR^d} \int_{(0, \infty)} e^{-t d_S(\vz)^2} d\mu_g(t) d\vz \\
        &= \int_{(0, \infty)} G_S(t) d\mu_g(t).
    \end{aligned}
    \]
    Now let $X, Y \in (\sR^d)^n$ satisfy $X \preceq Y$. By Lemma~\ref{lem:gaussian_comparison}, 
    \[
    G_X(t) \le G_Y(t)
    \]
    for every $t > 0$. Hence,
    \[
    \gD_{\phi,d}(Y) - \gD_{\phi,d}(X) = \int_{(0, \infty)} ( G_Y(t) - G_X(t) ) d\mu_g(t) \ge 0.
    \]
    This proves weak distance monotonicity.

    \paragraph{Equality at every Gaussian scale.}
    Assume now that $g$ has accumulating Laplace support and that $\gD_{\phi,d}(X) = \gD_{\phi,d}(Y)$. Define
    \[
    h(t) := G_Y(t) - G_X(t).
    \]
    Then $h(t) \ge 0$ for every $t>0$, while 
    \[
    \int_{(0, \infty)} h(t) d\mu_g(t) = 0.
    \]

    By Lemma~\ref{lem:gaussian_functional_analytic}, $h = G_Y - G_X$ is real analytic on $(0, \infty)$, and hence continuous. 
    
    We claim that $h(t) = 0$ for every $t \in \supp(\mu_g) \cap (0, \infty)$. Indeed, if $h(t_0) > 0$ at some such point $t_0$, continuity of $h$ gives an open neighborhood $I \subset (0, \infty)$ of $t_0$ and a constant $c > 0$ such that
    \[
    h(t) \ge c
    \]
    for every $t \in I$. Since $t_0 \in \supp(\mu_g)$, $\mu_g(I) > 0$, and consequently 
    \[
    \int_{(0, \infty)} h(t) d\mu_g(t) \ge \int_I h(t) d\mu_g(t) \ge c\mu_g(I) > 0,
    \]
    which is a contradiction. Thus, $h$ vanishes on $\supp(\mu_g) \cap (0, \infty)$. Since this set has an accumulation point in $(0, \infty)$, the identity theorem for real-analytic functions gives
    \[
    G_X(t) = G_Y(t)
    \]
    for every $t > 0$.

    \paragraph{Equality of equal-radius union volumes.}
    For a finite configuration $S$, define the equal-radius union-volume function
    \[
    U_S(r) := \Vol_d \left( \bigcup_{\vs \in \supp(S)} B_d(\vs, r) \right), \quad r \ge 0.
    \]
    For every $t > 0$,
    \[
    e^{-t d_S(\vz)^2} = \int_{d_S(\vz)^2}^\infty te^{-ts} ds = \int_0^\infty te^{-ts} \1_{d_S(\vz)^2 \le s} ds.
    \]
    Tonelli's theorem therefore gives
    \[
    \begin{aligned}
        G_S(t) &= \int_{\sR^d} e^{-t d_S(\vz)^2} d\vz \\
        &= t \int_0^{\infty} e^{-ts} U_S(\sqrt s) ds.
    \end{aligned}
    \]
    Since $G_X(t) = G_Y(t)$ for every $t > 0$,
    \[
    \int_0^{\infty} e^{-ts} \left( U_Y(\sqrt s) - U_X(\sqrt{s}) \right) ds = 0
    \]
    for every $t > 0$. Define
    \[
    q(s) := U_Y(\sqrt s) - U_X(\sqrt s).
    \]
    The function $q$ is measurable and locally integrable. 
    Moreover,
    \[
    |q(s)| \le U_X(\sqrt s) + U_Y(\sqrt s) \le 2nv_d s^{d/2},
    \]
    so $q$ has at most polynomial growth.
    By the standard uniqueness theorem for Laplace transforms of locally integrable functions of exponential order, $q(s) = 0$ for almost every $s > 0$. Equivalently,
    \[
    U_X(r) = U_Y(r)
    \]
    for almost every $r > 0$.

    \paragraph{Strictness.}
    Suppose $X$ and $Y$ are not congruent. Since $X \preceq Y$, $Y$ is a noncongruent expansion of $X$. By the large-radius strictness theorem of \citet[Theorem~1]{gorbovickis_strict_2013}, there exists $r_0 > 0$ such that
    \[
    U_X(r) < U_Y(r)
    \]
    for every $r > r_0$. Consequently,
    \[
    q(s) = U_Y(\sqrt s) - U_X(\sqrt s) > 0
    \]
    for every $s > r_0^2$, contradicting $q(s) = 0$ for almost every $s>0$.
    
    Conversely, if $X \cong Y$, Euclidean isometry invariance from Theorem~\ref{thm:basic_properties} gives
    \[
    \gD_{\phi,d}(X) = \gD_{\phi,d}(Y).
    \]
    Hence,
    \[
    \gD_{\phi,d}(X) = \gD_{\phi,d}(Y) \iff X \cong Y.
    \]
\end{proof}

\subsection{Kneser-Poulsen Connection}
\label{app:kneser_poulsen}

For a nonempty finite configuration $X$, define its equal-radius union-volume function by
\[
U_X(r) := \Vol_d \! \left( \bigcup_{\vx \in \supp(X)} B_d(\vx, r) \right), \quad r \ge 0.
\]
The equal-radius Kneser-Poulsen conjecture asserts that an expansive rearrangement of the centers of congruent Euclidean balls cannot decrease the volume of their union; in our notation,
\[
X \preceq Y \Longrightarrow U_X(r) \le U_Y(r) \text{ for every } r \ge 0.
\]
This is elementary in dimension $d=1$ and follows in dimension $d=2$ from the planar Kneser-Poulsen theorem of \citet{bezdek2002pushing}.
It remains open in general for $d \ge 3$.

We first express GKD directly in terms of these union volumes.

\begin{lemma}[Equal-radius union-volume representation] 
\label{lem:layer_cake}

Let $d \ge 1$, let $\phi$ be a normalized $d$-admissible radial kernel, and let $X \in (\sR^d)^n$ be nonempty.
Then
\[
\gD_{\phi,d}(X) = \int_{(0, \phi(0)]} U_X(\phi^{-1}(t)) \, dt.
\]
\end{lemma}
\begin{proof}[Proof of Lemma~\ref{lem:layer_cake}]
    Since $\phi$ is nonnegative, strictly decreasing, and integrable with respect to $r^{d-1} dr$, we have
    \[
    \lim_{r \to \infty} \phi(r) = 0.
    \]
    Hence
    \[
    \phi^{-1}: (0, \phi(0)] \to [0, \infty)
    \]
    is well-defined.

    For every $\vz \in \sR^d$,
    \[
    \phi(d_X(\vz)) = \int_{(0, \phi(0)]} \1_{t \le \phi(d_X(\vz))} \, dt.
    \]
    Tonelli's theorem gives
    \[
    \gD_{\phi,d}(X) = \int_{(0, \phi(0)]} \int_{\sR^d} \1_{t \le \phi(d_X(\vz))} \, d\vz \, dt.
    \]
    Because $\phi$ is strictly decreasing,
    \[
    t \le \phi(d_X(\vz)) \iff d_X(\vz) \le \phi^{-1}(t).
    \]
    Therefore,
    \[
    \left\{ \vz \in \sR^d \;\middle|\; d_X(\vz) \le \phi^{-1}(t) \right\} = \bigcup_{\vx \in \supp(X)} B_d(\vx, \phi^{-1}(t)),
    \]
    whose volume is $U_X(\phi^{-1}(t))$. Substitution yields the claimed identity.
\end{proof}

\begin{corollary}[Kneser-Poulsen monotonicity] \label{cor:kneser_poulsen}
    Suppose the equal-radius Kneser-Poulsen inequality holds in dimension $d$.
    Let $\phi$ be a normalized $d$-admissible radial kernel and let $X, Y \in (\sR^d)^n$ satisfy $X \preceq Y$.
    Then
    \[
    \gD_{\phi,d}(X) \le \gD_{\phi,d}(Y).
    \]
    Moreover, 
    \[
    \gD_{\phi,d}(X) = \gD_{\phi,d}(Y) \iff X \cong Y.
    \]
    Consequently, if $X \prec Y$, then
    \[
    \gD_{\phi,d}(X) < \gD_{\phi,d}(Y).
    \]
    In particular, this conclusion holds unconditionally for $d \in \{1, 2\}$, and conditionally on the equal-radius Kneser-Poulsen conjecture for every $d \ge 3$.
\end{corollary}
\begin{proof}[Proof of Corollary~\ref{cor:kneser_poulsen}]
    By the assumed Kneser-Poulsen inequality,
    \[
    U_X(r) \le U_Y(r) \text{ for every } r \ge 0.
    \]
    Lemma~\ref{lem:layer_cake} therefore gives
    \[
    \gD_{\phi,d}(X) \le \gD_{\phi,d}(Y).
    \]
    Now suppose that $X \ncong Y$. Since $X \preceq Y$, $Y$ is a noncongruent expansion of $X$. By the large-radius strictness theorem of \citet{gorbovickis_strict_2013}, there exists $r_0 > 0$ such that
    \[
    U_X(r) < U_Y(r) \text{ for every } r > r_0.
    \]
    Since $\phi(r_0) > 0$, the interval $(0, \phi(r_0))$ has positive Lebesgue measure. For every $t$ in this interval,
    \[
    \phi^{-1}(t) > r_0,
    \]
    and hence 
    \[
    U_X(\phi^{-1}(t)) < U_Y(\phi^{-1}(t)).
    \]
    The layer-cake representation therefore yields
    \[
    \gD_{\phi,d}(X) < \gD_{\phi,d}(Y).
    \]

    Conversely, if $X \cong Y$, Euclidean-isometry invariance from Theorem~\ref{thm:basic_properties} gives
    \[
    \gD_{\phi,d}(X) = \gD_{\phi,d}(Y).
    \]
    This proves the equality characterization. Since $X \prec Y$ implies $X \ncong Y$, the strict inequality follows as well.

    Finally, the equal-radius Kneser-Poulsen inequality is elementary for $d=1$ and follows for $d=2$ from \citet{bezdek2002pushing}; for $d \ge 3$, the conclusion is conditional on the conjecture.
\end{proof}

\subsection{Mironov--Prokhorenkova Axioms on Euclidean Inputs}
\label{app:mironov_axioms}

We first recall the formal axioms of \citet{mironov2025measuring}.

\begin{definition}[Mironov--Prokhorenkova axioms]
    Fix $n \ge 1$.
    Let $\gM_n$ be the set of symmetric nonnegative $n \times n$ matrices $A = (a_{ij})$ with zero diagonal satisfying
    \[
    a_{ij} = 0 \Rightarrow a_{ik} = a_{jk} \text{ for all }i, j, k.
    \]
    Let $F: \gM_n\to \sR$ be a permutation-invariant diversity function.
    The axioms of \citet{mironov2025measuring} are:

    \begin{itemize}
        \item[\textbf{(M)}] \textbf{Monotonicity.}
        Let $A = (a_{ij})$ and $B = (b_{ij})$ belong to $\gM_n$.
        If $a_{ij} \ge b_{ij}$ for all $i, j$, and at least one of the inequalities is strict, then $F(A) > F(B)$.

        \item[\textbf{(U)}] \textbf{Uniqueness.}
        Let $A = (a_{ij})$ and $B = (b_{ij})$ belong to $\gM_n$ and satisfy
        \[
        \begin{aligned}
            a_{ij} &= b_{ij} & \text{for all } i>2, \, j>2, \\
            b_{1j} &= b_{2j} & \text{for all } j, \\
            a_{2j} &= b_{2j} & \text{for all } j>2, \\
            a_{1j} &> 0 & \text{for all } j>1.
        \end{aligned}
        \]
        Then $F(A) > F(B)$.
        
        \item[\textbf{(C)}] \textbf{Continuity.}
        The function $F$ is continuous on this domain with respect to the subspace topology inherited from $\sR^{n \times n}$.
    \end{itemize}
\end{definition}

For a labeled configuration $X=(\vx_1,\dots,\vx_n)\in(\sR^d)^n$, let
\[
(D_X)_{ij}:=\|\vx_i-\vx_j\|,
\]
and define the set of Euclidean-realizable $n$-point distance matrices by
\[
\gE_{n,d} := \left\{ D_X\in\sR^{n\times n} \;\middle|\; X\in(\sR^d)^n \right\},
\]
equipped with the subspace topology inherited from $\sR^{n\times n}$. We call $X$ a realization of $D\in \gE_{n,d}$ if $D_X=D$. 

Define
\[
\overline{\gD}_{\phi,d,n}(D_X) := \gD_{\phi,d}(X).
\]
This is well-defined because labeled Euclidean configurations with the same pairwise distance matrix are congruent, and GKD is invariant under Euclidean isometries.

We first establish the continuity fact needed to pass from point configurations to their distance matrices.

For two labeled configurations $X = (\vx_1, \dots, \vx_n)$ and $Y = (\vy_1, \dots, \vy_n)$ in $(\sR^d)^n$, write
\[
\|X-Y\|_{\mathrm{prod}} := \left( \sum_{i=1}^n \|\vx_i-\vy_i\|^2 \right)^{1/2}.
\]

\begin{lemma}[Alignment of convergent realizations] \label{lem:alignment}
    Let $D^{(k)}, D \in \gE_{n,d}$ satisfy
    \[
    D^{(k)} \longrightarrow D
    \]
    in $\sR^{n \times n}$. Let $X \in (\sR^d)^n$ be any realization of $D$, and for each $k$, let $Y^{(k)} \in (\sR^d)^n$ be any realization of $D^{(k)}$. Then there exist Euclidean isometries $T_k: \sR^d \to \sR^d$ such that
    \[
    T_k(Y^{(k)}) \longrightarrow X
    \]
    in $(\sR^d)^n$.
\end{lemma}
\begin{proof}
    Translate each configuration so that its first point is the origin. Define
    \[
    \widetilde X := (0, \vx_2-\vx_1, \dots, \vx_n-\vx_1)
    \]
    and
    \[
    \widetilde Y^{(k)} := (0, \vy_2^{(k)}-\vy_1^{(k)}, \dots, \vy_n^{(k)} - \vy_1^{(k)}).
    \]
    For every $i$,
    \[
    \|\vy_i^{(k)} - \vy_1^{(k)}\| = D^{(k)}_{1i} \longrightarrow D_{1i}.
    \]
    Therefore, the sequence $\widetilde Y^{(k)}$ is bounded in $(\sR^d)^n$. 
    We claim that
    \[
    \inf_{Q \in \orth(d)} \left\| Q \widetilde Y^{(k)} - \widetilde X \right\|_{\mathrm {prod}} \longrightarrow 0,
    \]
    where $\orth(d)$ is the orthogonal group and $Q$ acts on each point of the configuration.

    Suppose otherwise. Then there exist $\eps > 0$ and a subsequence, still indexed by $k$, such that
    \[
    \inf_{Q \in \orth(d)} \left\| Q \widetilde Y^{(k)} - \widetilde X \right\|_{\mathrm{prod}} \ge \eps
    \]
    for every $k$. Since the sequence is bounded in a finite-dimensional Euclidean space, it has a further subsequence converging to some
    \[
    Z = (0, \vz_2, \dots, \vz_n) \in (\sR^d)^n.
    \]
    For every $i, j$,
    \[
    \|\vz_i-\vz_j\| = \lim_{k \to \infty} D^{(k)}_{ij} = D_{ij}.
    \]
    Thus, $Z$ and $\widetilde X$ have the same pairwise distance matrix. Since their first points are both zero, the polarization identity gives
    \[
    \langle \vz_i, \vz_j \rangle = \frac{D_{1i}^2 + D_{1j}^2 - D_{ij}^2}{2} = \langle \vx_i - \vx_1, \vx_j - \vx_1 \rangle.
    \]
    Hence the two configurations have the same Gram matrix, so the correspondence
    \[
    \vz_i \longmapsto \vx_i - \vx_1
    \]
    extends to a linear isometry between their spans. 
    Extending this isometry to the orthogonal complements gives some $Q \in \orth(d)$ such that
    \[
    QZ = \widetilde X.
    \]
    Along the convergent subsequence,
    \[
    Q \widetilde Y^{(k)} \longrightarrow QZ = \widetilde X,
    \]
    contradicting the lower bound by $\eps$. This proves the claim.

    For every $k$, compactness of $\orth(d)$ implies that the infimum is attained by some $Q_k \in \orth(d)$. Therefore,
    \[
    Q_k \widetilde Y^{(k)} \longrightarrow \widetilde X.
    \]
    Undoing the initial translations, define
    \[
    T_k(\vz) := \vx_1 + Q_k(\vz - \vy_1^{(k)}).
    \]
    Each $T_k$ is a Euclidean isometry, and
    \[
    T_k(Y^{(k)}) \longrightarrow X.
    \]
\end{proof}

\begin{proof}[Proof of Corollary~\ref{cor:mironov_axioms}]
\leavevmode

\paragraph{Permutation invariance.}
Let $\pi \in S_n$, and let $P_\pi$ be its permutation matrix. For $X = (\vx_1, \dots, \vx_n)$, define
\[
\pi X := (\vx_{\pi(1)}, \dots, \vx_{\pi(n)}).
\]
Then,
\[
D_{\pi X} = P_\pi D_X P_\pi^T.
\]
By permutation invariance from Theorem~\ref{thm:basic_properties},
\[
\begin{aligned}
    \overline{\gD}_{\phi,d,n} \left(P_{\pi} D_X P_\pi^T \right)&= \gD_{\phi,d}(\pi X) \\
    &=\gD_{\phi,d}(X) = \overline{\gD}_{\phi,d,n}(D_X).
\end{aligned}
\]

\paragraph{Monotonicity.}
Let $D_X, D_Y \in \gE_{n,d}$ satisfy
\[
(D_X)_{ij} \le (D_Y)_{ij}
\]
for every $i, j$. Equivalently, $X \preceq Y$. Theorem~\ref{thm:distance_monotonicity} gives
\[
\overline{\gD}_{\phi,d,n}(D_X) = \gD_{\phi,d}(X) \le \gD_{\phi,d}(Y) = \overline{\gD}_{\phi,d,n}(D_Y).
\]
If at least one entry inequality is strict, then $X \prec Y$, and the same theorem gives
\[
\overline{\gD}_{\phi,d,n}(D_X) < \overline{\gD}_{\phi,d,n}(D_Y).
\]
This is exactly the monotonicity axiom restricted to $\gE_{n,d}$.

\paragraph{Uniqueness.}
Let
\[
A = (a_{ij}), B = (b_{ij}) \in \gE_{n,d}
\]
satisfy the hypotheses of the uniqueness axiom above.

Choose realizations $X = (\vx_1, \dots, \vx_n)$ and $Y = (\vy_1, \dots, \vy_n)$ of $A$ and $B$, respectively. 

The condition $b_{1j} = b_{2j}$ for all $j$ implies, by taking $j=1$, that $\|\vy_1 - \vy_2\| = b_{12} = b_{11} = 0$, so $\vy_1 = \vy_2$. Moreover, the subconfigurations $(\vx_2, \dots, \vx_n)$ and $(\vy_2, \dots, \vy_n)$ have the same labeled pairwise distances and are therefore congruent. Applying a Euclidean isometry to $X$, we may assume
\[
\vx_i = \vy_i \text{ for every } i \ge 2.
\]
Let $S := (\vx_2, \dots, \vx_n)$. Since $a_{1j} > 0$ for every $j > 1$, the point $\vx_1$ is distinct from every point in $\supp(S)$. Strict addition monotonicity from Theorem~\ref{thm:basic_properties} therefore gives
\[
\gD_{\phi,d}(S) < \gD_{\phi,d}(X).
\]
On the other hand, $\vy_1 = \vy_2$ and $(\vy_2, \dots, \vy_n) = S$, so duplicate invariance gives
\[
\gD_{\phi,d}(Y) = \gD_{\phi,d}(S).
\]
Consequently,
\[
\overline{\gD}_{\phi,d,n}(A) > \overline{\gD}_{\phi,d,n}(B),
\]
which is the uniqueness axiom restricted to $\gE_{n,d}$.

\paragraph{Continuity.}
Let $D^{(k)} \rightarrow D$ be a convergent sequence in $\gE_{n,d}$. Choose any realization $X \in (\sR^d)^n$ of $D$, and for every $k$, choose any realization $Y^{(k)} \in (\sR^d)^n$ of $D^{(k)}$.

By Lemma~\ref{lem:alignment}, there exist Euclidean isometries $T_k$ such that
\[
T_k(Y^{(k)}) \longrightarrow X.
\]
Using Euclidean isometry invariance and continuity from Theorem~\ref{thm:basic_properties}, 
\[
\begin{aligned}
    \overline{\gD}_{\phi,d,n}(D^{(k)}) &= \gD_{\phi,d}(Y^{(k)}) \\
    &= \gD_{\phi,d}(T_k(Y^{(k)})) \\
    &\longrightarrow \gD_{\phi,d}(X) = \overline{\gD}_{\phi,d,n}(D).
\end{aligned}
\]
Therefore, $\overline{\gD}_{\phi,d,n}: \gE_{n,d} \to \sR$ is continuous in the subspace topology on $\gE_{n,d}$.
\end{proof}

\section{Canonical Kernel and Exact Computation}
\label{app:computation}

\subsection{Canonical-Kernel Properties}
\label{app:canonical_kernel}

\begin{proof}[Proof of Lemma~\ref{lem:canonical_properties}]
    Fix $d \ge 1$ and $\sigma > 0$, and write
    \[
    \alpha := \frac{d}{2} + 1.
    \]
    The canonical kernel is
    \[
    \psi_{d,\sigma}(r) =
    \frac{\Gamma(\alpha)}{\pi^{d/2}\sigma^d} \left( 1 + \frac{r^2}{\sigma^2} \right)^{-\alpha}.
    \]
    It is continuous, nonnegative, and strictly decreasing on $[0, \infty)$.

    We first verify normalization. Using polar coordinates and the substitution
    \[
    u := \frac{r^2}{\sigma^2},
    \]
    we obtain
    \[
    \begin{aligned}
        \int_{\sR^d} \psi_{d,\sigma}(\|\vz\|) \, d\vz &=
        s_{d-1} \frac{\Gamma(\alpha)}{\pi^{d/2} \sigma^d} \int_0^{\infty} r^{d-1} \left( 1+\frac{r^2}{\sigma^2} \right)^{-\alpha} \, dr \\
        &= \frac{\Gamma(\alpha)}{\Gamma(d/2)} \int_0^{\infty} u^{d/2-1} (1+u)^{-\alpha} \, du
    \end{aligned}
    \]
    The remaining integral is
    \[
    B \left( \frac{d}{2}, 1 \right) = \frac{\Gamma(d/2)\Gamma(1)}{\Gamma(d/2+1)} = \frac{\Gamma(d/2)}{\Gamma(\alpha)}.
    \]
    Hence
    \[
    \int_{\sR^d} \psi_{d,\sigma}(\|\vz\|) \,d\vz = 1.
    \]
    Thus, $\psi_{d,\sigma}$ is a normalized $d$-admissible radial kernel.

    Now define
    \[
    g_{d,\sigma}(s) := \frac{\Gamma(\alpha)}{\pi^{d/2} \sigma^d} \left( 1+\frac{s}{\sigma^2} \right)^{-\alpha}, \quad s \ge 0.
    \]
    Then
    \[
    \psi_{d,\sigma}(r)=g_{d,\sigma}(r^2).
    \]
    For every integer $k \ge 0$,
    \[
    (-1)^k g_{d,\sigma}^{(k)} (s) =
    \frac{\Gamma(\alpha)}{\pi^{d/2}\sigma^d} \frac{1}{\sigma^{2k}} \left( 1 + \frac{s}{\sigma^2} \right)^{-\alpha-k} \prod_{i=0}^{k-1} (\alpha+i),
    \]
    where the product is interpreted as $1$ for $k=0$.
    Thus,
    \[
    (-1)^k g_{d,\sigma}^{(k)}(s) > 0
    \]
    for every $k \ge 0$ and $s \ge 0$, so $g_{d,\sigma}$ is completely monotone.
    
    To identify its Bernstein-Widder representing measure, use
    \[
    (s+\sigma^2)^{-\alpha} = \frac{1}{\Gamma(\alpha)} \int_0^{\infty} t^{\alpha-1} e^{-(s+\sigma^2)t} \, dt.
    \]
    Since
    \[
    \left( 1+\frac{s}{\sigma^2} \right)^{-\alpha} = \sigma^{2\alpha} (s+\sigma^2)^{-\alpha},
    \]
    we obtain
    \[
    \begin{aligned}
        g_{d,\sigma}(s) &= \frac{\Gamma(\alpha)}{\pi^{d/2} \sigma^d} \frac{\sigma^{2\alpha}}{\Gamma(\alpha)}  \int_0^{\infty} t^{\alpha-1} e^{-st}e^{-\sigma^2t} \, dt \\
        &= \frac{\sigma^2}{\pi^{d/2}} \int_0^{\infty} t^{d/2} e^{-st} e^{-\sigma^2t} \, dt.
    \end{aligned}
    \]
    Therefore, its representing measure has density
    \[
    d\mu_{d,\sigma}(t) = \frac{\sigma^2}{\pi^{d/2}} t^{d/2} e^{-\sigma^2 t} dt.
    \]
    This density is strictly positive for every $t > 0$, so $\supp(\mu_{d,\sigma})$ has accumulation points throughout $(0, \infty)$.
    Hence $g_{d,\sigma}$ has accumulating Laplace support.
\end{proof}

\begin{lemma}[Scaling identity] \label{lem:canonical_scaling}
    For every nonempty $X \in (\sR^d)^n$,
    \[
    \gD_{\psi_{d,\sigma},d}(X)
    =
    \gD_{\psi_{d,1},d}(X/\sigma),
    \]
    where $X / \sigma = (\vx_1/\sigma, \dots, \vx_n / \sigma)$.
\end{lemma}
\begin{proof}
    The canonical kernel satisfies
    \[
    \psi_{d,\sigma}(r) = \sigma^{-d} \psi_{d,1} (r/\sigma).
    \]
    Therefore,
    \[
    \gD_{\psi_{d,\sigma},d}(X) =
    \int_{\sR^d} \sigma^{-d} \psi_{d,1} \left( \frac{d_X(\vz)}{\sigma} \right) \, d\vz.
    \]
    Using the substitution $\vz = \sigma \vu$, we have $d\vz = \sigma^d \, d\vu$ and
    \[
    \frac{d_X(\sigma \vu)}{\sigma} =
    \min_{\vx \in \supp(X)} \left\| \vu - \frac{\vx}{\sigma} \right\| = d_{X/\sigma}(\vu).
    \]
    Consequently,
    \[
    \gD_{\psi_{d,\sigma},d}(X) =
    \int_{\sR^d} \psi_{d,1} (d_{X/\sigma}(\vu)) \, d\vu = 
    \gD_{\psi_{d,1},d}(X/\sigma).
    \]
\end{proof}

\begin{lemma}[Dimensional marginal consistency]
\label{lem:canonical_marginal}
    Let $g \ge d \ge 1$, $\sigma > 0$, and $r \ge 0$. Then
    \[
    \int_{\sR^{g-d}} \psi_{g,\sigma} \left( \sqrt{r^2+\|\vv\|^2} \right) \, d\vv = \psi_{d,\sigma}(r).
    \]
\end{lemma}
\begin{proof}
    Let $(\rvu, \rvv) \in \sR^d \times \sR^{g-d}$ have density
    \[
    (\vu, \vv) \longmapsto
    \psi_{g,\sigma} \left( \sqrt{\|\vu\|^2 + \|\vv\|^2} \right).
    \]
    By Section~\ref{sec:canonical_kernel}, this is the density of
    \[
    t_g \left( \vzero, \frac{\sigma^2}{2} I_{g}, 2 \right).
    \]
    By the standard marginal-closure property of the multivariate Student-$t$ distribution \citep{ding2016conditional}, the first $d$ coordinates satisfy
    \[
    \rvu \sim t_d \left( \vzero, \frac{\sigma^2}{2} I_d, 2 \right) .
    \]
    Their density is therefore
    \[
    \vu \longmapsto \psi_{d,\sigma}(\|\vu\|).
    \]
    On the other hand, by the definition of a marginal density, it is
    \[
    \vu \longmapsto \int_{\sR^{g-d}} \psi_{g,\sigma} \left( \sqrt{\|\vu\|^2 + \|\vv\|^2} \right) \, d\vv.
    \]
    Equating these expressions and choosing any $\vu$ with $\|\vu\| = r$ gives the claimed identity.
\end{proof}

\begin{proof}[Proof of Proposition~\ref{prop:canonical_ambient_dimension}]
    Let $T: \sR^d \to \sR^{g}$ be an affine isometric embedding.
    After composing $T$ with a Euclidean isometry of $\sR^{g}$, which does not change canonical GKD, we may assume without loss of generality that
    \[
    T(\vu) = (\vu, \vzero) \in \sR^d \times \sR^{g-d}.
    \]
    For $(\vu, \vv) \in \sR^d \times \sR^{g-d}$,
    \[
    \begin{aligned}
        d_{T(X)}((\vu, \vv)) &= \min_{\vx \in \supp(X)} \sqrt{\|\vu-\vx\|^2 + \|\vv\|^2} \\
        &= \sqrt{d_X(\vu)^2 + \|\vv\|^2}.
    \end{aligned}
    \]
    Therefore, Tonelli's theorem and Lemma~\ref{lem:canonical_marginal} give
    \[ 
    \begin{aligned}
        \gD_{\psi_{g,\sigma},g}(T(X)) &= \int_{\sR^d} \int_{\sR^{g-d}} \psi_{g,\sigma} \left( \sqrt{d_X(\vu)^2 + \|\vv\|^2} \right) \, d\vv \, d\vu \\
        &= \int_{\sR^d} \psi_{d,\sigma}(d_X(\vu)) \, d\vu \\
        &= \gD_{\psi_{d,\sigma},d}(X).
    \end{aligned}
    \]
\end{proof}

\begin{proposition}[Pairwise canonical GKD]
\label{prop:pairwise_diversity}
    Let $d \ge 1$, $\sigma>0$, and $\vx,\vy\in\sR^d$.
    If $r=\|\vx-\vy\|$, then
    \[
    \gD_{\psi_{d,\sigma},d}(\vx,\vy)
    =
    1+\frac{r}{\sqrt{r^2+4\sigma^2}}.
    \]
\end{proposition}
\begin{proof}
    Let $r := \|\vx-\vy\|$. By Euclidean isometry invariance, we may assume
    \[
    \vx = -\frac{r}{2} \ve_1, \quad \vy = \frac{r}{2} \ve_1.
    \]
    Their perpendicular bisector is the hyperplane $z_1 = 0$.
    The point $\vx$ is the nearest on the half-space $z_1 \le 0$, while $\vy$ is nearest on the half-space $z_1 \ge 0$. By symmetry,
    \[
    \gD_{\psi_{d,\sigma},d}(\vx, \vy) = 2 \int_{\{z_1 \le 0\}} \psi_{d,\sigma} ( \|\vz-\vx\|) \, d\vz.
    \]
    Translating by $\vx$ gives
    \[
    \gD_{\psi_{d,\sigma},d}(\vx, \vy) = 2 \int_{\{u_1 \le r/2\}} \psi_{d,\sigma}(\|\vu\|) \, d\vu.
    \]
    Let $\rvu$ have density
    \[
    \vu \longmapsto \psi_{d,\sigma}(\|\vu\|).
    \]
    By Lemma~\ref{lem:canonical_marginal}, its first coordinate has density
    \[
    p_\sigma(t) = \frac{1}{2\sigma} \left(1 + \frac{t^2}{\sigma^2} \right)^{-3/2}.
    \]
    Hence 
    \[
    \gD_{\psi_{d,\sigma},d}(\vx, \vy) = 2 \Pr \left[ \ru_1 \le \frac{r}{2} \right].
    \]
    An antiderivative of $p_\sigma$ is
    \[
    \frac{t}{2\sqrt{t^2+\sigma^2}},
    \]
    and therefore 
    \[
    \Pr[\ru_1 \le t] = \frac{1}{2} + \frac{t}{2\sqrt{t^2 + \sigma^2}}.
    \]
    Substituting $t = r/2$ yields
    \[
    \begin{aligned}
        \gD_{\psi_{d,\sigma},d}(\vx, \vy) &= 1 + \frac{r}{\sqrt{r^2+4\sigma^2}}.
    \end{aligned}
    \]
    In particular,
    \[
    \gD_{\psi_{d,\sigma},d}(\vx, \vy) = \frac 32 \iff r = \frac{2}{\sqrt 3} \sigma.
    \]
\end{proof}

\paragraph{Why two degrees of freedom?}

Ambient-dimensional consistency alone does not single out the choice $\nu=2$. For any $\nu > 0$, define
\[
\psi_{d,\sigma}^{(\nu)} (r) := 
\frac{\Gamma((d+\nu)/2)}{\Gamma(\nu/2) \pi^{d/2} \sigma^d} \left( 1 + \frac{r^2}{\sigma^2} \right)^{-(d+\nu)/2}.
\]
Then
\[
\psi_{d,\sigma}^{(2)} = \psi_{d,\sigma},
\]
and $\vu \longmapsto \psi_{d,\sigma}^{(\nu)}(\|\vu\|)$ is the density of
\[
t_d \left( \vzero, \frac{\sigma^2}{\nu} I_d, \nu \right).
\]
For every fixed $\nu$, this family has the same dimensional marginal-consistency property.
Two additional simplifications used by our algorithms occur specifically when $\nu = 2$. 

\paragraph{Uniform-ball transformation.}
Define
\[
T(\vu) := \frac{\vu}{\sqrt{\sigma^2 + \|\vu\|^2}}.
\]
This is a smooth bijection from $\sR^d$ onto the unit ball $B_d(\vzero, 1)$, with inverse
\[
T^{-1}(\vy) = \frac{\sigma \vy}{\sqrt{1 - \|\vy\|^2}}.
\]

\begin{lemma}[Jacobian of the radial transformation]
\label{lem:radial_transform_jacobian}
    For
    \[
    T(\vu) := \frac{\vu}{\sqrt{\sigma^2 + \|\vu\|^2}},
    \]
    \[
    \left| \det DT(\vu) \right| =
    \frac{1}{\sigma^d} \left(1 + \frac{\|\vu\|^2}{\sigma^2} \right)^{-(d/2+1)}.
    \]
    Equivalently, for $\vy = T(\vu)$,
    \[
    \left| \det DT^{-1}(\vy) \right| =
    \sigma^d (1-\|\vy\|^2)^{-(d/2+1)}. 
    \]
\end{lemma}
\begin{proof}
    For $a, b \in \{1, \dots, d\}$,
    \[
    \frac{\partial T_a}{\partial u_b} = 
    (\sigma^2 + \|\vu\|^2)^{-1/2} \delta_{ab} - (\sigma^2 + \|\vu\|^2)^{-3/2} u_a u_b.
    \]
    Hence
    \[
    DT(\vu) = (\sigma^2 + \|\vu\|^2)^{-1/2} \left(I - \frac{\vu \vu^T}{\sigma^2 + \|\vu\|^2} \right).
    \]
    By the matrix determinant lemma,
    \[
    \det \left( I - \frac{\vu \vu^T}{\sigma^2 + \|\vu\|^2} \right) =
    1 - \frac{\|\vu\|^2}{\sigma^2 + \|\vu\|^2}  =
    \frac{\sigma^2}{\sigma^2 + \|\vu\|^2}.
    \]
    Therefore,
    \[
    \left| \det DT(\vu) \right| = 
    \sigma^2 (\sigma^2 + \|\vu\|^2)^{-(d/2+1)} =
    \frac{1}{\sigma^d} \left(1 + \frac{\|\vu\|^2}{\sigma^2} \right)^{-(d/2+1)}.
    \]
    Since
    \[
    1 - \|\vy\|^2 = \frac{\sigma^2}{\sigma^2 + \|\vu\|^2},
    \]
    taking reciprocals gives
    \[
    \left| \det DT^{-1}(\vy) \right| = \sigma^d (1-\|\vy\|^2)^{-(d/2+1)}.
    \]
\end{proof}

Let $\rvu$ have density $\psi_{d,\sigma}^{(\nu)} (\|\vu\|)$ and set $\rvy = T(\rvu)$. Since 
\[
1 + \frac{\|\vu\|^2}{\sigma^2} = \frac{1}{1-\|\vy\|^2},
\]
Lemma~\ref{lem:radial_transform_jacobian} and the change of variables formula give
\[
f_{\rvy}(\vy) = \frac{\Gamma((d+\nu)/2)}{\Gamma(\nu/2)\pi^{d/2}} (1-\|\vy\|^2)^{\nu/2-1}, \quad \|\vy\|<1.
\]
This is the spherical beta distribution associated with the Student-$t$ radial transformation \citep{jones2013beta}.
When $\nu = 2$, the exponent is zero and
\[
f_{\rvy}(\vy) = \frac{\Gamma(d/2+1)}{\pi^{d/2}} = \frac{1}{\Vol_d(B_d(\vzero, 1))}.
\]
Hence
\[
\rvy \sim \Unif(B_d(\vzero, 1)).
\]
This is the uniform-ball representation used by the exact sampling oracle in Appendix~\ref{app:canonical_oracles}.

\paragraph{Divergence simplification.}
After scaling to $\sigma = 1$, define
\[
\fF_\nu(\vu) := \frac{\vu}{(1+\|\vu\|^2)^{(d+\nu-2)/2}}.
\]
Writing
\[
p := \frac{d+\nu-2}{2},
\]
a direct differentiation gives
\[
\begin{aligned}
    \nabla \! \cdot \fF_\nu(\vu) &=
    d(1+\|\vu\|^2)^{-p} - 2p\|\vu\|^2 (1+\|\vu\|^2)^{-p-1} \\
    &= \left( d+(2-\nu)\|\vu\|^2 \right) (1+\|\vu\|^2)^{-(d+\nu)/2}.
\end{aligned}
\]
Thus, $\nu = 2$ is exactly the value at which the quadratic term disappears:
\[
\nabla \! \cdot \left( \frac{\vu}{(1+\|\vu\|^2)^{d/2}} \right)
= \frac{d}{(1+\|\vu\|^2)^{d/2+1}}.
\]
Since
\[
c_d := \frac{\Gamma(d/2)}{2 \pi^{d/2}}
\]
satisfies
\[
c_d d = \frac{\Gamma(d/2+1)}{\pi^{d/2}},
\]
we obtain
\[
c_d \, \nabla \cdot \left( \frac{\vu}{(1+\|\vu\|^2)^{d/2}} \right) =
\psi_{d,1}(\|\vu\|).
\]
This is precisely the divergence identity used in Appendix~\ref{app:voronoi_divergence}.

\subsection{Exact Computation in One Dimension} \label{app:comp_1d}

\begin{proof}[Proof of Theorem~\ref{thm:comp_1d}]
    Sort the input points so that
    \[
    x_1 \le x_2 \le \dots \le x_n,
    \]
    and define $\Delta_i := x_{i+1} - x_i$ for $i \in \{1, \dots, n-1\}$.
    Repeated points cause no difficulty: if $x_i = x_{i+1}$, then $\Delta_i = 0$ and the corresponding term in the formula vanishes.

    For $d=1$, the canonical kernel is
    \[
    \psi_{1,\sigma}(r) = \frac{1}{2\sigma} \left(1 + \frac{r^2}{\sigma^2} \right)^{-3/2}.
    \]
    An antiderivative on $[0, \infty)$ is
    \[
    F_\sigma(r) :=  \frac{r}{2\sqrt{r^2+\sigma^2}}, \quad F_\sigma'(r) = \psi_{1,\sigma}(r).
    \]
    Up to shared endpoints of measure zero, $\sR$ is decomposed into the two unbounded intervals
    \[
    (-\infty, x_1], \quad [x_n, \infty)
    \]
    and the intervals $[x_i, x_{i+1}]$ for $i \in \{1, \dots, n-1\}$.
    Intervals corresponding to duplicate consecutive points are degenerate and contribute zero.

    On $(-\infty, x_1)$, the point $x_1$ is a nearest input point, so 
    \[
    \int_{-\infty}^{x_1} \psi_{1,\sigma}(d_X(z)) \, dz = \int_0^{\infty} \psi_{1,\sigma}(r) \, dr = \frac{1}{2}.
    \]
    Similarly,
    \[
    \int_{x_n}^{\infty} \psi_{1,\sigma}(d_X(z)) \, dz = \frac{1}{2}.
    \]
    Thus, the two unbounded intervals contribute a total of $1$.

    Now fix $i \in \{1, \dots, n-1\}$. On $[x_i, x_{i+1}]$, the nearest input point is $x_i$ to the left of the midpoint and $x_{i+1}$ to the right.
    Therefore, by symmetry,
    \[
    \begin{aligned}
        \int_{x_i}^{x_{i+1}} \psi_{1,\sigma}(d_X(z)) \, dz &= 2\int_0^{\Delta_i/2} \psi_{1,\sigma}(r) \, dr \\
        &= 2 \left[\frac{r}{2\sqrt{r^2+\sigma^2}} \right]_0^{\Delta_i/2} \\
        &= \frac{\Delta_i}{\sqrt{\Delta_i^2 + 4\sigma^2}}.
    \end{aligned}
    \]
    Summing all contributions gives
    \[
    \gD_{\psi_{1,\sigma},1}(X) = 1 + \sum_{i=1}^{n-1} \frac{\Delta_i}{\sqrt{\Delta_i^2 + 4\sigma^2}}.
    \]
    For $n = 1$, the sum is empty and the same formula gives $\gD_{\psi_{1,\sigma},1}(X) = 1$.
    
    Sorting takes $O(n \log n)$ time. Once the points are sorted, evaluating the $n-1$ terms requires $O(n)$ elementary real-RAM operations.
    Hence $\gD_{\psi_{1,\sigma},1}(X)$ can be computed in $O(n \log n)$ time.
\end{proof}

\subsection{Voronoi-Divergence Reduction} \label{app:voronoi_divergence}

By Lemma~\ref{lem:canonical_scaling}, it suffices to consider the standard canonical kernel $\psi_{d,1}$.

After removing duplicate points, write $X = (\vx_1, \dots, \vx_m)$ for the pairwise distinct points. If $m=1$, then $\gD_{\psi_{d,1},d}(X) = 1$, so assume $m \ge 2$.

For each $i \in \{1, \dots, m\}$, let
\[
P_i := \left\{ \vz \in \sR^d \;\middle|\; \|\vz-\vx_i\| \le \|\vz-\vx_j\| \text{ for every } j \right\}
\]
be the Voronoi cell of $\vx_i$.

For adjacent cells $P_i$ and $P_j$, meaning cells whose intersection contains a $(d-1)$-dimensional face, let
\[
F_{ij} := P_i \cap P_j, \quad d_{ij} := \|\vx_i-\vx_j\|, \quad \vm_{ij} := \frac{\vx_i+\vx_j}{2},
\]
and let 
\[
\vn_{ij} := \frac{\vx_j-\vx_i}{d_{ij}}, \quad \vn_{ji} = -\vn_{ij}.
\]
The face $F_{ij}$ lies in the perpendicular-bisector hyperplane
\[
H_{ij}:=\vm_{ij}+\vn_{ij}^{\perp}.
\]
Choose a linear isometry
\[
L_{ij} : \sR^{d-1} \to \vn_{ij}^\perp
\]
and define
\[
Q_{ij} := \left\{ \vu \in \sR^{d-1} \;\middle|\; \vm_{ij} + L_{ij} \vu \in F_{ij} \right\}.
\]
Thus, $Q_{ij}$ is the common Voronoi face expressed in coordinates centered at the midpoint of $\vx_i$ and $\vx_j$.

\begin{lemma}[Voronoi-divergence reduction] \label{lem:voronoi_divergence}
    For $d \in \{2, 3\}$,
    \[
    \gD_{\psi_{d,1},d}(X) = 1 + c_d \sum_{\{i, j\}} d_{ij} \int_{Q_{ij}} \frac{d\vu}{\left(1 + d_{ij}^2/4 + \|\vu\|^2 \right)^{d/2}},
    \]
    where
    \[
    c_d := \frac{\Gamma(d/2)}{2\pi^{d/2}},
    \]
    and the sum ranges over unordered adjacent pairs $\{i, j\}$.
\end{lemma}
\begin{proof}
    Since the Voronoi cells cover $\sR^d$ and their boundaries have $d$-dimensional Lebesgue measure zero, 
    \[
    \gD_{\psi_{d,1},d}(X) = \sum_{i=1}^m \int_{P_i} \psi_{d,1}(\|\vz-\vx_i\|) \, d\vz.
    \]
    Define
    \[
    \fF_i(\vz) := c_d \frac{\vz-\vx_i}{\left(1+\|\vz-\vx_i\|^2 \right)^{d/2}}.
    \]
    The divergence identity from Appendix~\ref{app:canonical_kernel}, translated by $\vx_i$, gives
    \[
    \nabla \! \cdot \fF_i(\vz) = \psi_{d,1}(\|\vz-\vx_i\|).
    \]
    For sufficiently large $R > 0$, let $B_R := B_d(\vzero, R)$, so that all points of $X$ lie in its interior.
    Each $P_i \cap B_R$ is a bounded convex region whose boundary is piecewise smooth, so the divergence theorem applies and gives
    \[
    \int_{P_i \cap B_R} \psi_{d,1} (\|\vz-\vx_i\|) \, d\vz = \int_{\partial (P_i \cap B_R)} \fF_i(\vz) \cdot \vn \, dS.
    \]
    Summing over $i$ yields
    \[
    \int_{B_R} \psi_{d,1}(d_X(\vz)) \, d\vz = S_R + \sum_{\{i,j\}} C_{ij}(R),
    \]
    where the sum is over unordered pairs of adjacent Voronoi cells, meaning pairs whose intersection contains a $(d-1)$-dimensional face.
    
    The outer-boundary contribution is
    \[
    S_R := \sum_{i=1}^m \int_{P_i \cap \partial B_R} \fF_i(\vz) \cdot \frac{\vz}{R} \, dS
    \]
    and the contribution of a common Voronoi face is
    \[
    C_{ij}(R) := \int_{F_{ij} \cap B_R} \left( \fF_i(\vz) \cdot \vn_{ij} + \fF_j(\vz) \cdot \vn_{ji} \right) dS.
    \]
    
    No general-position assumption is required here.
    Intersections of three or more distinct Voronoi cells have dimension at most $d-2$. Indeed, for distinct $i, j, k$,
    \[
    P_i \cap P_j \cap P_k \subseteq H_{ij} \cap H_{jk},
    \]
    where $H_{ij}$ and $H_{jk}$ are distinct perpendicular-bisector hyperplanes.
    If they were the same, then the hyperplane would reflect $x_j$ to both $x_i$ and $x_k$, contradicting $x_i \neq x_k$.
    Their intersection is therefore either empty or has dimension at most $d-2$. Such sets have zero $(d-1)$-dimensional measure and contribute no flux.
    
    \paragraph{Outer boundary.}
    
    Write $\vz = R\vw$ where $\|\vw\| = 1$. Then,
    \[
    \begin{aligned}
    R^{d-1} \fF_i(R\vw) \cdot \vw &=
    c_d \frac{R^{d-1} (R-\vx_i \cdot \vw)}{\left(1 + \|R\vw-\vx_i\|^2 \right)^{d/2}} \\
    &= c_d \frac{1-\vx_i \cdot \vw/R}{\left( 1 - \frac{2\vx_i\cdot\vw}{R} + \frac{1 + \|\vx_i\|^2}{R^2} \right)^{d/2}}.
    \end{aligned}
    \]
    Because the set of points is finite,
    \[
    R^{d-1} \fF_i(R \vw) \cdot \vw \longrightarrow c_d
    \]
    uniformly in both $i$ and $\vw$ as $R \to \infty$.
    
    The sets $P_i \cap \partial B_R$ partition $\partial B_R$ up to a set of surface measure zero. Under $\vz = R\vw$,
    \[
    dS = R^{d-1} \, d\omega,
    \]
    where $d\omega$ is the surface measure on the unit sphere. 
    Hence
    \[
    \lim_{R \to \infty} S_R = c_d s_{d-1} = 1.
    \]
    
    \paragraph{Voronoi-face contribution.}
    
    For every $\vz \in F_{ij}$, $\|\vz-\vx_i\| = \|\vz-\vx_j\|$. Therefore,
    \[
    \begin{aligned}
        \fF_i(\vz) \cdot \vn_{ij} + \fF_j(\vz) \cdot \vn_{ji} &= c_d \left( \frac{\vz-\vx_i}{(1+\|\vz-\vx_i\|^2)^{d/2}} - \frac{\vz-\vx_j}{(1+\|\vz-\vx_j\|^2)^{d/2}} \right) \cdot \vn_{ij} \\
        &= c_d \frac{d_{ij}}{(1+\|\vz-\vx_i\|^2)^{d/2}}.
    \end{aligned}
    \]
    Thus,
    \[
    C_{ij}(R) = c_d d_{ij} \int_{F_{ij} \cap B_R} \frac{dS}{(1+\|\vz-\vx_i\|^2)^{d/2}}.
    \]    
    For
    \[
    \vz = \vm_{ij} + L_{ij} \vu,
    \]
    we have
    \[
    \vm_{ij} - \vx_i = \frac{d_{ij}}{2} \vn_{ij},
    \]
    which is orthogonal to $L_{ij} \vu$. Therefore
    \[
    \|\vz-\vx_i\|^2 = \frac{d_{ij}^2}{4} + \|\vu\|^2.
    \]
    As $R \to \infty$, the sets $F_{ij} \cap B_R$ increase to $F_{ij}$. Since the integrand is nonnegative and $L_{ij}$ preserves $(d-1)$-dimensional measure, the monotone convergence theorem gives
    \[
    \lim_{R \to \infty} C_{ij}(R) = c_d d_{ij} \int_{Q_{ij}} \frac{d\vu}{\left( 1 + d_{ij}^2/4 + \|\vu\|^2\right)^{d/2}}.
    \]
    Finally,
    \[
    \int_{B_R} \psi_{d,1}(d_X(\vz)) \, d\vz \longrightarrow \gD_{\psi_{d,1},d}(X)
    \]
    by monotone convergence. There are finitely many adjacent pairs, so we may pass to the limit term-by-term in the face sum. Combining the face limits with
    \[
    \lim_{R \to \infty} S_R = 1
    \]
    gives
    \[
    \gD_{\psi_{d,1},d}(X) = 1 + c_d \sum_{\{i,j\}} d_{ij} \int_{Q_{ij}} \frac{d\vu}{\left( 1 + d_{ij}^2/4 + \|\vu\|^2 \right)^{d/2}},
    \]
    which proves Lemma~\ref{lem:voronoi_divergence}.
\end{proof}

\subsection{Exact Computation in Two Dimensions} \label{app:comp_2d}

\begin{proof}[Proof of Theorem~\ref{thm:comp_2d}]
    By Lemma~\ref{lem:canonical_scaling}, after rescaling the input it suffices to assume $\sigma = 1$. By duplicate invariance, we may also remove duplicate points.

    Let $X = (\vx_1, \dots, \vx_m)$ be the resulting pairwise distinct points. If $m = 1$, then the GKD value is $1$, so assume $m \ge 2$.
    
    Construct the planar Voronoi diagram of $X$. For each unordered adjacent pair $\{i, j\}$, let $E_{ij} := P_i \cap P_j$ be their common Voronoi edge and set $d_{ij} := \|\vx_i - \vx_j\|$.
    
    Since $c_2 = \frac{1}{2\pi}$, Lemma~\ref{lem:voronoi_divergence} gives
    \[
    \gD_{\psi_{2,1},2}(X) = 1 + \frac{1}{2\pi} \sum_{\{i,j\}} d_{ij} \int_{Q_{ij}} \frac{dt}{1+d_{ij}^2/4+t^2}.
    \]
    Because a Voronoi edge is a convex subset of its supporting line, $Q_{ij}$ is an interval with ordered endpoints
    \[
    -\infty \le a_{ij} < b_{ij} \le +\infty
    \]
    where the interval may be bounded, a ray, or the entire line. Define
    \[
    A_{ij} := \sqrt{1+\frac{d_{ij}^2}{4}}.
    \]
    Then,
    \[
    \int_{Q_{ij}} \frac{dt}{1+d_{ij}^2/4+t^2}
    = \int_{a_{ij}}^{b_{ij}} \frac{dt}{A_{ij}^2+t^2} 
    = \frac{1}{A_{ij}} \left[ \arctan\left( \frac{t}{A_{ij}} \right) \right]_{a_{ij}}^{b_{ij}}
    \]
    where
    \[
    \arctan(+\infty) = \frac{\pi}{2}, \quad \arctan(-\infty) = -\frac{\pi}{2}.
    \]
    Hence the contribution of $E_{ij}$ is 
    \[
    \frac{d_{ij}}{2\pi A_{ij}} \left( \arctan\left( \frac{b_{ij}}{A_{ij}} \right) - \arctan \left( \frac{a_{ij}}{A_{ij}} \right) \right).
    \]
    Therefore,
    \[
    \gD_{\psi_{2,1},2}(X) =
    1+\sum_{\{i,j\}} \frac{d_{ij}}{2\pi A_{ij}} \left( \arctan\left( \frac{b_{ij}}{A_{ij}} \right) - \arctan\left( \frac{a_{ij}}{A_{ij}} \right) \right).
    \]
    A planar Voronoi diagram of $m$ points has $O(m)$ total combinatorial complexity and can be constructed in $O(m \log m)$ time \citep{fortune_sweepline_1987}.
    No general-position assumption is needed.
    Cocircular points may create Voronoi vertices incident to more than three cells, but the one-dimensional Voronoi edges are still segments or rays.
    If all points are collinear, the nonempty Voronoi edges are complete lines, which are handled by taking $a_{ij} = -\infty$ and $b_{ij} = +\infty$.

    Once the diagram is constructed, its geometric representation provides the endpoints or directions of every Voronoi edge.
    Relative to the coordinate system defining $Q_{ij}$ in Appendix~\ref{app:voronoi_divergence}, the corresponding finite endpoints or ray directions can be converted to  $a_{ij}$ and $b_{ij}$ in $O(1)$ elementary real-RAM operations per edge.
    Each edge contribution then requires only $O(1)$ additional operations, so evaluating all contributions takes $O(m) = O(n)$ time.

    Including the duplicate removal and the $O(n)$ rescaling from Lemma~\ref{lem:canonical_scaling}, the total running time is $O(n \log n)$ for every $\sigma > 0$.
    This proves Theorem~\ref{thm:comp_2d}.
\end{proof}

\subsection{Planar Face Integration} \label{app:planar_face_integration}

\begin{lemma}[Planar face integration]
\label{lem:planar_face_integration}
    Let $Q \subseteq \sR^2$ be a nonempty convex polygonal region with nonempty interior, possibly unbounded, whose boundary consists of $k$ line segments, rays, or lines.
    For every $A > 0$,
    \[
    I_A(Q) := \int_Q \frac{d \vu}{(A^2 + \|\vu\|^2)^{3/2}}
    \]
    can be evaluated using $O(k+1)$ elementary real-RAM operations.
\end{lemma}
\begin{proof}
For bounded $Q$, the quantity $A I_A(Q)$ is the solid angle subtended by $Q$, embedded in a plane, from a point at perpendicular distance $A$.
Thus, the bounded case could alternatively be evaluated by triangulating $Q$ and using classical closed-form formulas for the solid angle of a triangle \citep{van1983solid}. We instead give a boundary formula that also handles unbounded regions.

Define
\[
\fG_A(\vu) := \frac{\vu}{A\sqrt{A^2+\|\vu\|^2} \left(A+\sqrt{A^2+\|\vu\|^2} \right)}.
\]
Let
\[
r := \|\vu\|, \quad s := \sqrt{A^2 + r^2},
\]
and write
\[
\fG_A(\vu) = \vu g(r), \quad g(r) := \frac{1}{As(A+s)}.
\]
For a radial scalar function $g$ in two dimensions,
\[
\nabla \! \cdot (\vu g(r)) = 2g(r) + rg'(r).
\]
Since
\[
g'(r) = -\frac{r(2s+A)}{As^3(A+s)^2} =
-\frac{r(2s+A)}{s^2(A+s)} g(r),
\]
we obtain
\[
\begin{aligned}
    \nabla \! \cdot \fG_A(\vu)
    &= g(r) \left( 2 - \frac{r^2(2s+A)}{s^2(A+s)} \right) \\
    &= g(r) \left(2 - \frac{(s-A)(2s+A)}{s^2} \right) \\
    &= g(r) \frac{A(A+s)}{s^2} \\
    &= \frac{1}{s^3} = \frac{1}{(A^2 + \|\vu\|^2)^{3/2}}.
\end{aligned}
\]
For $R > 0$, let
\[
Q_R := Q \cap B_2(\vzero, R).
\]
For all sufficiently large $R$, $Q_R$ is a nonempty bounded convex region with piecewise smooth boundary. The divergence theorem gives
\[
\int_{Q_R} \frac{d\vu}{(A^2 + \|\vu\|^2)^{3/2}} = \int_{\partial Q_R} \fG_A(\vu) \cdot \vn(\vu) \, ds.
\]
Up to finitely many intersection points, $\partial Q_R$ consists of portions of the straight boundary pieces of $Q$ together with circular arcs on $\partial B_2(\vzero, R)$. 

\paragraph{Straight boundary pieces.}
Let $E$ be a straight boundary piece of $Q$, which may be a segment, ray, or complete line. Let $\vn_E$ be its outward unit normal and choose a unit tangent vector $\vtau_E$ perpendicular to $\vn_E$. The supporting line of $E$ has an equation
\[
\vn_E \cdot \vu = c_E
\]
for some constant $c_E \in \sR$, so we may parameterize
\[
\vu_E(t) = c_E \vn_E + t \vtau_E, \quad t \in J_E,
\]
where $J_E$ is an interval whose endpoints $a_E, b_E$ lie in $\sR \cup \{\pm \infty\}$.

Since
\[
\|\vu_E(t)\|^2 = c_E^2 + t^2, \quad 
\vu_E(t) \cdot \vn_E = c_E,
\]
we have 
\[
\fG_A(\vu_E(t)) \cdot \vn_E = \frac{c_E}{A \sqrt{A^2 + c_E^2 + t^2} \left( A + \sqrt{A^2+c_E^2+t^2} \right)}.
\]
If $c_E = 0$, this flux vanishes identically. For $c_E \neq 0$, define
\[
\Phi_{A,c_E}(t) := \frac{c_E}{A|c_E|} \left[ \arctan \left( \frac{t}{|c_E|} \right) - \arctan \left( \frac{At}{|c_E|\sqrt{A^2+c_E^2+t^2}} \right) \right].
\]
Direct differentiation gives
\[
\Phi_{A,c_E}'(t) = \fG_A(\vu_E(t)) \cdot \vn_E.
\]
For all sufficiently large $R$, the portion of $E$ contained in $\partial Q_R$ is parameterized by 
\[
J_{E,R} = J_E \cap \left[ -\sqrt{R^2-c_E^2}, \sqrt{R^2-c_E^2} \right].
\]
Let $a_{E,R}$ and $b_{E,R}$ denote its endpoints. As $R \to \infty$,
\[
a_{E,R} \to a_E, \quad b_{E,R} \to b_E
\]
in the extended real line, so the limiting flux contribution of $E$ is
\[
\Phi_{A,c_E}(b_{E}) - \Phi_{A,c_E}(a_{E}).
\]
For infinite endpoints, 
\[
\Phi_{A,c_E}(+\infty) = \frac{c_E}{A|c_E|} \left( \frac{\pi}{2} - \arctan \frac{A}{|c_E|} \right), \quad \Phi_{A,c_E}(-\infty) = -\Phi_{A,c_E}(+\infty).
\]
Thus, each straight boundary piece contributes a constant-size elementary expression.

\paragraph{Boundary at infinity.}
Write
\[
\vw(\theta) := (\cos \theta, \sin \theta)
\]
and define
\[
\Theta_R := \left\{ \theta\in[0, 2\pi) \mid R\vw(\theta) \in Q \right\}.
\]
On $\partial B_2(\vzero, R)$, the outward normal of $Q_R$ along its circular boundary is $\vw(\theta)$ and $ds = R \, d\theta$. Hence the total circular flux equals
\[
\frac{R^2}{A \sqrt{A^2+R^2} \left(A + \sqrt{A^2+R^2} \right)} |\Theta_R|.
\]
The scalar factor converges to $1/A$ as $R \to \infty$.

Define the recession cone of $Q$ by
\[
\rec(Q) := \left\{ \vw \in \sR^2 \mid \vu + \lambda \vw \in Q \text{ for every } \vu \in Q, \ \lambda \ge 0 \right\}
\]
and let
\[
\beta(Q) := | \{ \theta \in [0, 2\pi) \mid \vw(\theta) \in \rec(Q) \} |.
\]
We claim that
\[
|\Theta_R| \longrightarrow \beta(Q).
\]
Write $Q$ as the intersection of finitely many half-planes,
\[
Q = \bigcap_{i = 1}^p \left\{ \vu \in \sR^2 \; \middle|\; \vn_i \cdot \vu \le h_i \right\}.
\]
Then
\[
R \vw(\theta) \in Q \iff \vn_i \cdot \vw(\theta) \le \frac{h_i}{R} \text{ for every } i.
\]
For every direction satisfying
\[
\vn_i \cdot \vw(\theta) \neq 0 \text{ for every }i,
\]
these inequalities eventually have the same truth values as
\[
\vn_i \cdot \vw(\theta) \le 0 \text{ for every } i,
\]
which is precisely the condition $\vw(\theta) \in \rec(Q)$. 
Only finitely many directions satisfy $\vn_i \cdot \vw(\theta) = 0$ for some $i$, so the exceptional set has angular measure zero. Thus,
\[
\1_{\theta \in \Theta_R} \longrightarrow \1_{\vw(\theta) \in \rec(Q)}
\]
for almost every $\theta$. Dominated convergence yields
\[
|\Theta_R| \longrightarrow \beta(Q).
\]
Consequently, the limiting circular-boundary contribution is
\[
\frac{\beta(Q)}{A}.
\]
Since $Q_R$ increases to $Q$ and the integrand is nonnegative, the monotone convergence theorem gives
\[
I_A(Q_R) \longrightarrow I_A(Q).
\]
Combining the straight-boundary and circular-boundary limits yields
\[
I_A(Q) = \sum_{E \in \gE(Q)} \left(\Phi_{A,c_E}(b_E) - \Phi_{A,c_E}(a_E) \right) + \frac{\beta(Q)}{A},
\]
where $\gE(Q)$ denotes the set of straight boundary pieces of $Q$, and the summand is interpreted as zero when $c_E = 0$.

It remains to compute $\beta(Q)$. From the polygonal boundary representation, the possibilities are as follows:

\begin{itemize}
    \item If $Q$ is bounded, then $\beta(Q) = 0$.
    \item If $Q = \sR^2$, then $\beta(Q) = 2\pi$.
    \item If $Q$ is a half-plane, then $\beta(Q) = \pi$.
    \item If $Q$ is a strip between two parallel lines, then $\rec(Q)$ is a line and $\beta(Q) = 0$.
    \item In every remaining unbounded case, the unbounded polygonal boundary has two rays.
    Let $\vr_1$ and $\vr_2$ be unit vectors in their recession directions. Then,
    \[
    \rec(Q) = \left\{ \lambda_1 \vr_1 + \lambda_2 \vr_2 \mid \lambda_1, \lambda_2 \ge 0 \right\},
    \]
    and $\beta(Q) = \arccos(\vr_1 \cdot \vr_2)$.
\end{itemize}

These cases and the corresponding recession directions can be identified in one scan of the $k$ boundary pieces.
Thus, $\beta(Q)$ can be computed in $O(k+1)$ elementary real-RAM operations.
Each straight boundary contribution requires $O(1)$ operations, so the entire integral $I_A(Q)$ can be evaluated using $O(k+1)$ operations.
\end{proof}

\subsection{Exact Computation in Three Dimensions}
\label{app:comp_3d}

\begin{proof}[Proof of Theorem~\ref{thm:comp_3d}]
    By Lemma~\ref{lem:canonical_scaling}, after rescaling the input it suffices to assume $\sigma=1$.
    By duplicate invariance, we may also remove duplicate points.

    If only one distinct point remains, then the GKD value is $1$.
    Hence assume that
    \[
    X = (\vx_1, \dots, \vx_m)
    \]
    contains $m \ge 2$ pairwise distinct points. 

    Construct the three-dimensional Voronoi diagram of $X$.
    For every unordered adjacent pair $\{i, j\}$, let $F_{ij} := P_i \cap P_j$ denote their common two-dimensional Voronoi face and set $d_{ij} := \|\vx_i - \vx_j\|$.
    Since
    \[
    c_3 = \frac{\Gamma(3/2)}{2\pi^{3/2}} = \frac{1}{4\pi},
    \]
    Lemma~\ref{lem:voronoi_divergence} gives
    \[
    \gD_{\psi_{3,1},3}(X) = 1 + \frac{1}{4\pi} \sum_{\{i,j\}} d_{ij} \int_{Q_{ij}} \frac{d\vu}{\left(1 + d_{ij}^2/4 + \|\vu\|^2 \right)^{3/2}},
    \]
    where $Q_{ij} \subseteq \sR^2$ is the planar representation of $F_{ij}$ introduced in Appendix~\ref{app:voronoi_divergence}.
    
    For each face, define
    \[
    A_{ij} := \sqrt{1+\frac{d_{ij}^2}{4}}.
    \]
    Then its integral is
    \[
    I_{A_{ij}}(Q_{ij}) = \int_{Q_{ij}} \frac{d\vu}{(A_{ij}^2 + \|\vu\|^2)^{3/2}}.
    \]
    Relative to the planar coordinate system defining $Q_{ij}$ in Appendix~\ref{app:voronoi_divergence}, the vertices and recession directions of $F_{ij}$ can be converted to the corresponding line segments, rays, or lines of $Q_{ij}$ using $O(k_{ij}+1)$ elementary real-RAM operations.
    Hence, after constructing the Voronoi diagram, the total coordinate-conversion and integral cost is 
    \[
    O \left( |\gF| + \sum_{F \in \gF} k(F) \right),
    \]
    where $\gF$ denotes the set of two-dimensional Voronoi faces and $k(F)$ is the number of one-dimensional boundary pieces of $F$.

    We now bound the incidence term
    \[
    \sum_{F \in \gF} k(F).
    \]
    For each Voronoi cell $P_i$, let $e(P_i)$ denote the number of its one-dimensional boundary pieces, including bounded edges, rays, and lines.
    Since $P_i$ is the intersection of at most $m-1 \le n-1$ half-spaces, it has at most $n-1$ two-dimensional faces.

    We claim that $e(P_i) = O(n)$ for every $i$.
    If $P_i$ is bounded, this follows from the standard planar-graph bound for the boundary graph of a three-dimensional convex polytope.
    If $P_i$ is unbounded, truncate it by a sufficiently large cube $C$ centered at the origin, chosen so that all bounded vertices and edges of $P_i$ lie in the interior of $C$ and no one-dimensional boundary piece of $P_i$ is contained in $\partial C$.
    Such a choice is possible because only finitely many cube sizes violate the latter condition.
    The polytope
    \[
    P_i' := P_i \cap C
    \]
    has at most $n-1+6 = n+5$ faces, and every one-dimensional boundary piece of $P_i$ is preserved as an edge of $P_i'$, possibly after truncation. Since a three-dimensional convex polytope with $f$ faces has at most $3f-6$ edges, $e(P_i) \le e(P_i') = O(n)$.

    For a fixed cell $P_i$, every one-dimensional piece is incident to exactly two of its two-dimensional faces. 
    Therefore,
    \[
    \sum_{\substack{F \in \gF \\ F \text{ incident to } P_i}} k(F) = 2e(P_i).
    \]
    Every two-dimensional Voronoi face is incident to exactly two Voronoi cells.
    Summing the preceding identity over all cells thus gives
    \[
    2 \sum_{F \in \gF} k(F) = 2\sum_{i=1}^m e(P_i),
    \]
    and hence
    \[
    \sum_{F \in \gF} k(F) = \sum_{i=1}^m e(P_i)= O(n^2).
    \]
    Also,
    \[
    |\gF| \le \binom{m}{2} = O(n^2),
    \]
    since a pair of convex Voronoi cells shares at most one two-dimensional face. Consequently,
    \[
    |\gF| + \sum_{F \in \gF} k(F) \in O(n^2),
    \]
    so all face integrals can be evaluated in $O(n^2)$ elementary real-RAM operations once the Voronoi diagram is available.
    
    A three-dimensional Voronoi diagram can be constructed in $O(n^2)$ time in the real-RAM model \citep{chazelle_optimal_1993}.
    Therefore, including Voronoi construction, duplicate removal, and the initial rescaling from Lemma~\ref{lem:canonical_scaling}, the total running time is $O(n^2)$ for every $\sigma > 0$.
\end{proof}

\section{Randomized Approximation}\label{app:estimator}

\subsection{The Self-Adjusting Union-Volume Estimator} \label{sec:self_coverage_estimator}

We use the self-adjusting coverage estimator of  \citet{karp1989approx}.
Their algorithm and analysis apply to unions of finite sets.
\citet{bringmann_2010} adapted the same approach to unions of measurable geometric bodies using volume, sampling, and membership oracles, and further allowed these oracles to be approximate.
For completeness, we state the exact-oracle version needed here, retaining the arbitrary failure probability $\delta$ from the analysis of \citet{karp1989approx}.

Let $(\Omega, \Sigma, \mu)$ be a measure space, and let $B_1, \dots, B_m \in \Sigma$ satisfy $0 < \mu(B_i) < \infty$ for every $i \in \{1, \dots, m\}$. Define
\[
V := \mu \left( \bigcup_{i=1}^m B_i \right), \quad W := \sum_{i=1}^m \mu(B_i).
\]
Assume access to the following operations:
\begin{enumerate}
    \item compute $\mu(B_i)$;
    \item sample a point $\ry$ from $B_i$ according to
    \[
    \Pr[\ry \in A] = \frac{\mu(A \cap B_i)}{\mu(B_i)}
    \]
    for every measurable $A \subseteq \Omega$;
    \item given $y \in \Omega$ and $i \in \{1, \dots, m\}$, determine whether $y \in B_i$.
\end{enumerate}
A trial first chooses an index $I \in \{1, \dots, m\}$ according to
\[
\Pr[I = i] = \frac{\mu(B_i)}{W}
\]
and then samples $\ry$ from the normalized restriction of $\mu$ to $B_I$.
It next repeatedly chooses independent indices
\[
J \sim \Unif(\{1, \dots, m\})
\]
and tests whether $\ry \in B_J$, ending the trial at the first successful query.

For $\eps, \delta \in (0, 1)$, define the membership-query budget
\[
B := \left\lceil \frac{8(1+\eps)m \log(2/\delta)}{\eps^2} \right\rceil.
\]
Run consecutive trials and stop after the $B$-th membership query.
Let $N_B$ denote the number of trials completed within these $B$ queries; an incomplete final trial is not counted.
Define
\[
\widehat V := \begin{cases}
    \dfrac{BW}{mN_B}, & (N_B > 0), \\[1ex]
    W, & (N_B = 0).
\end{cases}
\]

\begin{lemma}[Self-adjusting union-volume estimator] \label{lem:self_adjusting_union}
    Under the assumptions above,
    \[
    \Pr \left[ (1-\eps)V \le \widehat V \le (1+\eps)V \right] \ge 1 - \delta.
    \]
\end{lemma}
\begin{proof}
    For $y \in \Omega$, let
    \[
    c(y) := \sum_{i=1}^m \1_{y \in B_i},
    \]
    and for $k \in \{1, \dots, m\}$ define
    \[
    R_k := \{y \in \Omega \mid c(y) = k\}, \quad r_k := \mu(R_k).
    \]
    Since the solids $B_i$ are measurable, so are $c$ and the sets $R_k$. Also, from the definition above,
    \[
    V = \sum_{k=1}^m r_k, \quad W = \int_{\Omega} c(y) d\mu(y) = \sum_{k=1}^m kr_k.
    \]
    Moreover, the point $\ry$ sampled at the beginning of a trial satisfies
    \[
    \Pr[\ry \in R_k]
    = \sum_{i=1}^m \frac{\mu(B_i)}{W} \frac{\mu(R_k \cap B_i)}{\mu(B_i)}
    = \frac{kr_k}{W}.
    \]
    Conditional on $\ry \in R_k$, each membership query succeeds independently with probability $k/m$, and hence the trial length satisfies
    \[
    L \mid (\ry \in R_k) \sim \Geom \left( \frac{k}{m} \right)
    \]
    where $\Geom(p)$ counts the number of membership queries through the first success.
    
    These are exactly the coverage classes and conditional trial-length distributions used by \citet{karp1989approx}. 
    The same replacement of finite cardinalities by volumes is made explicitly by \citet{bringmann_2010} in their continuous-body formulation.
    Thus, the self-adjusting coverage analysis applies to the present exact-oracle setting.
    In particular, Theorem I of \citet{karp1989approx} gives the stated fixed-budget guarantee.
    
    For the ceiling in our definition of $B$, define
    \[
    \delta_B := 2\exp \left( -\frac{\eps^2 B}{8(1+\eps)m} \right).
    \]
    By the definition of $B$, we have $\delta_B \le \delta$, and
    \[
    B = \frac{8(1+\eps) m \log(2/\delta_B)}{\eps^2}.
    \]
    Applying Theorem I of \citet{karp1989approx} with failure probability $\delta_B$ therefore gives
    \[
    \Pr \left[ N_B > 0 \text{ and } (1-\eps)V \le \frac{BW}{mN_B} \le (1+\eps)V \right] \ge 1 - \delta_B \ge 1-\delta.
    \]
    On this event, $\widehat V = BW/mN_B$, which proves the claimed probability bound.
    The fallback value for $N_B = 0$ is used only outside this success event and therefore does not affect the guarantee.
\end{proof}

\subsection{Sampling and Membership Oracles for Canonical Solids}
\label{app:canonical_oracles}

For $X = (\vx_1, \dots, \vx_n) \in (\sR^d)^n$ and $\sigma > 0$, define
\[
S_i := S_{\psi_{d,\sigma},d}(\vx_i)
= \left\{ (\vz,h) \in \sR^d \times \sR_{\ge 0} \;\middle|\; 0 \le h \le \psi_{d,\sigma} (\|\vz-\vx_i\|) \right\}.
\]
We now construct the volume, membership, and uniform-sampling oracles required by Lemma~\ref{lem:self_adjusting_union}. 

\begin{lemma}[Canonical-solid oracles] \label{lem:canonical_oracles}
Under the randomized elementary real-RAM model of Section~\ref{sec:randomized_estimator}, each canonical solid $S_i$ supports the following operations:
\begin{enumerate}
    \item its volume can be returned in $O(1)$ time;
    \item membership of a point $(\vz, h) \in \sR^{d+1}$ can be tested in $O(d)$ time;
    \item a point can be sampled exactly from the normalized Lebesgue measure on $S_i$ in $O(d)$ time.
\end{enumerate}
\end{lemma}

\begin{proof}
    By normalization of the canonical kernel,
    \[
    \Vol_{d+1}(S_i) = 1,
    \]
    so the volume oracle is immediate.

    For membership,
    \[
    (\vz, h) \in S_i \iff 0 \le h \le \psi_{d,\sigma}(\|\vz-\vx_i\|).
    \]
    Computing $\|\vz-\vx_i\|^2$ takes $O(d)$ arithmetic operations. Since $d$ is an integer, the integer or half-integer powers and normalization constant in $\psi_{d,\sigma}$ can also be evaluated in $O(d)$ elementary real-RAM operations. Thus, membership can be tested in $O(d)$ time.

    It remains to construct an exact uniform sampler.

    \paragraph{Sampling the under-graph solid.} 
    First consider the solid centered at the origin,
    \[
    S_0 = S_{\psi_{d,\sigma},d}(\vzero).
    \]
    Suppose $\vz$ is sampled from the probability density
    \[
    p(\vz) = \psi_{d,\sigma}(\|\vz\|),
    \]
    and, conditional on $\vz$, independently sample
    \[
    H \sim \Unif \! \left( 0, \psi_{d,\sigma}(\|\vz\|) \right).
    \]
    The joint density of $(\vz, H)$ on $S_0$ is
    \[
    \psi_{d,\sigma}(\|\vz\|) \cdot \frac{1}{\psi_{d,\sigma}(\|\vz\|)} = 1.
    \]
    Since $\Vol_{d+1}(S_0) = 1$, this is exactly the uniform probability distribution on $S_0$. Thus, it remains only to sample $\vz$ from the canonical density.

    \paragraph{Sampling the canonical density.}
    Recall from Appendix~\ref{app:canonical_kernel} the radial map
    \[
    T(\vu) = \frac{\vu}{\sqrt{\sigma^2+\|\vu\|^2}}, \quad T^{-1}(\vy) = \frac{\sigma\vy}{\sqrt{1-\|\vy\|^2}}
    \]
    By Lemma~\ref{lem:radial_transform_jacobian},
    \[
    \left| \det DT(\vu) \right| = \frac{1}{\sigma^d} \left( 1 + \frac{\|\vu\|^2}{\sigma^2} \right)^{-(d/2+1)}.
    \]
    Since
    \[
    \psi_{d,\sigma}(\|\vu\|) = \frac{\Gamma(d/2+1)}{\pi^{d/2}} \left| \det DT(\vu) \right| = \frac{1}{v_d} \left| \det DT(\vu) \right|,
    \]
    where
    \[
    v_d = \Vol_d(B_d(\vzero, 1)) = \frac{\pi^{d/2}}{\Gamma(d/2+1)},
    \]
    the change of variables formula shows that $T$ maps the canonical density to the uniform probability distribution on the unit ball.
    Consequently, if
    \[
    \vy \sim \Unif(B_d(\vzero, 1))
    \]
    then
    \[
    \vz = \frac{\sigma \vy}{\sqrt{1 - \|\vy\|^2}}
    \]
    has density $\psi_{d,\sigma}(\|\vz\|)$.

    To sample $\vy$ exactly, draw $G_1, \dots, G_{d+2} \overset{\mathrm{i.i.d}}{\sim} \gN(0, 1)$ and set
    \[
    \vy = \frac{(G_1, \dots, G_d)}{\sqrt{G_1^2 + \dots + G_{d+2}^2}}.
    \]
    Normalizing a standard Gaussian vector gives a uniform point on $S^{d+1}$ \citep{muller1959sphere}, and projecting onto its first $d$ coordinates gives the uniform distribution on $B_d(\vzero, 1)$ \citep{mijatovic2018projections}.

    Therefore, an exact sampler for $S_i$ is:

    \begin{enumerate}
        \item sample $G_1, \dots, G_{d+2} \overset{\mathrm{i.i.d}}{\sim} \gN(0, 1)$ and set
        \[
        \vy = \frac{(G_1, \dots, G_d)}{\sqrt{G_1^2 + \dots + G_{d+2}^2}};
        \]
        \item set 
        \[
        \vz = \frac{\sigma \vy}{\sqrt{1-\|\vy\|^2}};
        \]
        \item independently sample $U \sim \Unif(0, 1)$ and set
        \[
        H := U \psi_{d,\sigma}(\|\vz\|);
        \]
        \item return $(\vx_i + \vz, H)$.
    \end{enumerate}

    The denominator in the second step is positive almost surely because the boundary of the unit ball has probability zero. 

    Generating and processing the $d+2$ Gaussian variables, applying $T^{-1}$, evaluating the kernel, sampling $U$, and translating by $\vx_i$ require $O(d)$ elementary real-RAM operations.
    Hence the complete sampler runs in $O(d)$ time.
\end{proof}

\subsection{Proof of the Randomized Approximation Theorem}
\label{app:estimator_proof}

\begin{proof}[Proof of Theorem~\ref{thm:estimator}]
For each $i \in \{1, \dots, n\}$, let
\[
S_i := S_{\psi_{d,\sigma},d}(\vx_i).
\]
By Proposition~\ref{prop:under_graph},
\[
\gD_{\psi_{d,\sigma},d}(X) =
\Vol_{d+1} \left( \bigcup_{i=1}^n S_i \right),
\]
and each $S_i$ has volume $1$.
By Lemma~\ref{lem:canonical_oracles}, these solids provide the exact volume, uniform-sampling, and membership oracles required by Lemma~\ref{lem:self_adjusting_union}.

Apply that lemma to 
\[
(\Omega, \Sigma, \mu) = \left( \sR^{d+1}, \gB(\sR^{d+1}), \Vol_{d+1} \right)
\]
with $B_i = S_i$. In its notation,
\[
m = n, \quad W = \sum_{i=1}^n \Vol_{d+1}(S_i) = n, \quad V = \gD_{\psi_{d,\sigma},d}(X).
\]
Hence its estimator specializes to
\[
\widehat V = \begin{cases}
    \dfrac{B}{N_B}, & (N_B > 0), \\[1ex]
    n, & (N_B = 0),
\end{cases}
\]
which is exactly the estimator $\widehat D$ defined in Section~\ref{sec:randomized_estimator}. Therefore, Lemma~\ref{lem:self_adjusting_union} gives
\[
\Pr\left[ \left| \widehat D - \gD_{\psi_{d,\sigma},d}(X) \right| \le \eps \, \gD_{\psi_{d,\sigma},d}(X) \right] \ge 1-\delta.
\]
It remains to bound the running time.
By Lemma~\ref{lem:canonical_oracles}, sampling uniformly from a canonical solid and performing a membership query take $O(d)$ time.
Because all solids have volume $1$, the initial index of each trial is uniform on $\{1, \dots, n\}$.
The membership-query indices are also uniform by construction.
These indices can be sampled in $O(1)$ time in the randomized real-RAM model.

The algorithm performs exactly $B$ membership queries.
Since every initiated trial performs at least one membership query, at most $B$ trials are initiated, and hence at most $B$ uniform-solid samples are generated.
Thus, the total running time is
\[
O(Bd) = O \left( \frac{nd}{\eps^2} \log \frac{2}{\delta} \right),
\]
which proves the theorem.
\end{proof}

\section{Additional Axiom-Stress-Test Details}
\label{app:experiments}

\subsection{Comparison Measures and Implementation}
\label{app:exp_measures}

\paragraph{Comparison measures.}
For a configuration $X=(x_1,\ldots,x_n)$, we compare GKD with average pairwise distance, the Vendi Score \citep{friedman2023vendi}, determinant-based DPP diversity \citep{mironov2025measuring, wilhelm2018practical}, and R\'enyi kernel entropy (RKE) \citep{jalali2023information}.

For the similarity-based measures, we use the RBF similarity matrix
\[
K_{ij} = \exp\!\left( -\frac{\|x_i-x_j\|^2}{2h^2} \right).
\]
Average pairwise distance is
\[
D_{\mathrm{APD}}(X) = \frac{2}{n(n-1)} \sum_{i<j}\|x_i-x_j\|.
\]
The determinant-based DPP score is
\[
D_{\mathrm{DPP}}(X) = \det K.
\]
Because $K_{ii}=1$, we have $\operatorname{tr}K=n$.
Let
$\lambda_1,\ldots,\lambda_n$ denote the eigenvalues of $K/n$. 
The Vendi Score is
\[
D_{\mathrm{Vendi}}(X) = \exp\!\left(-\sum_{i=1}^n \lambda_i\log \lambda_i\right),
\]
with the convention $0\log0=0$.
The order-two R\'enyi kernel entropy (RKE) is
\[
D_{\mathrm{RKE}}(X) =
- \log \left( \frac{1}{n^2} \sum_{i,j} K_{ij}^2 \right) = 
- \log \operatorname{tr} \! \left( \left( \frac{K}{n} \right)^2 \right).
\]
All three similarity-based measures use the same RBF bandwidth within a given experiment.

\paragraph{Implementation and reproducibility.}
All experimental code is implemented in Python.
Exact software dependencies, package versions, and the computing environment used for the final reproducibility audit are provided in the supplementary material.
Our empirical results concern diversity values and statistical comparison frequencies rather than wall-clock performance, and we do not make hardware-dependent runtime comparisons.

\subsection{Controlled Euclidean Stress Tests}
\label{app:exp_controlled}

\begin{figure}[t]
\centering
\includegraphics[width=\linewidth]{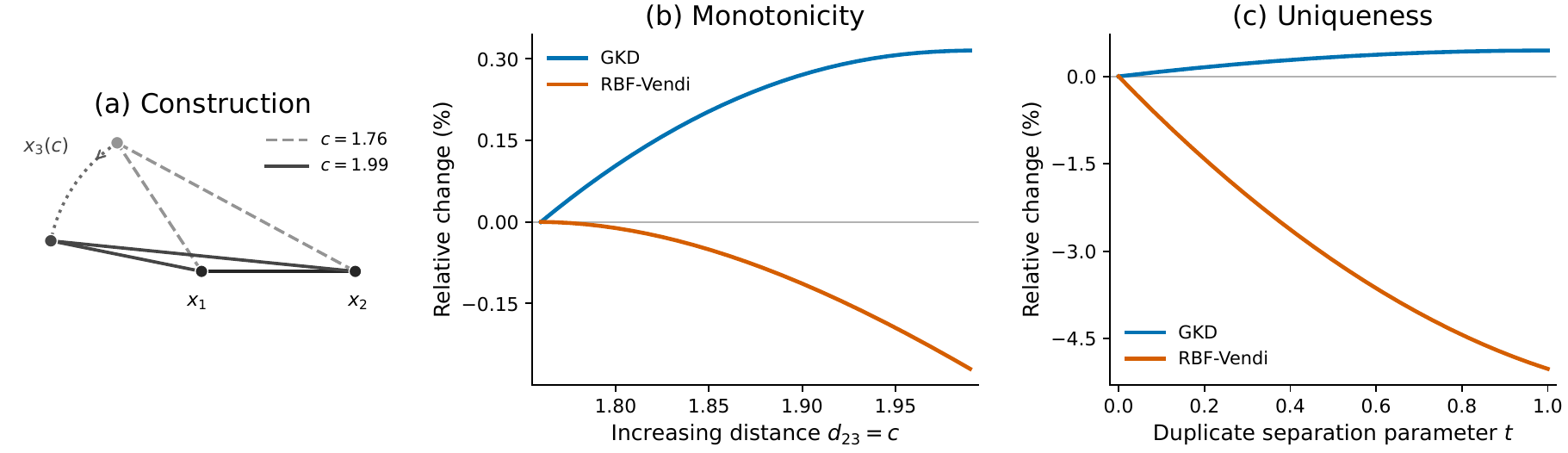}
\caption{
\textbf{Controlled axiom stress tests.}
\textbf{(a)} Three-point monotonicity construction with
$d_{12}=d_{13}=1$ and $d_{23}=c$, where $c$ increases from
$1.76$ to $1.99$.
\textbf{(b)} GKD increases along this expansion, while the Vendi Score decreases.
\textbf{(c)} In the uniqueness construction, separating one member of a duplicate pair increases GKD while the Vendi Score decreases.
The behavior of all comparison measures is reported in Table~\ref{tab:controlled_numeric}.
}
\label{fig:axiom_stress}
\end{figure}

This appendix gives the explicit constructions and numerical endpoint values for the controlled examples in Section~\ref{sec:exp_axiom_stress}.
These examples isolate the monotonicity and uniqueness axioms in simple low-dimensional Euclidean configurations. Figure~\ref{fig:axiom_stress} illustrates the monotonicity construction and plots the GKD and Vendi Score behavior for both examples.

Throughout this subsection, GKD uses the canonical kernel with $\sigma=1$, and the similarity-based measures use bandwidth $h=1$.
GKD is evaluated exactly, so these comparisons contain no Monte Carlo error.

\paragraph{Distance monotonicity.}
We use the three-point planar family
\[
x_1=(0,0), \quad x_2=(1,0),
\]
and
\[
x_3(c) = \left( 1-\frac{c^2}{2}, \sqrt{1-\left(1-\frac{c^2}{2}\right)^2} \right), \quad c\in[1.76,1.99].
\]
A direct calculation gives
\[
\|x_1-x_2\| = \|x_1-x_3(c)\| = 1, \quad \|x_2-x_3(c)\| = c.
\]
Thus, increasing $c$ changes exactly one pairwise distance while leaving the other two unchanged.
Any diversity measure satisfying strict distance monotonicity must therefore increase along this path.

At the endpoints $c = 1.76$ and $c = 1.99$, exact GKD increases from $1.88847$ to $1.89442$.
In contrast, the Vendi Score decreases from $2.29378$ to $2.28758$, and the determinant DPP score decreases from $0.37543$ to $0.34676$.
Average pairwise distance and RKE increase along this particular path.

\paragraph{Uniqueness.}
For the uniqueness example, we use the one-dimensional family
\[
X(t) = \left( 0,\, 0,\, \frac12,\, \frac12-\frac{t}{4} \right), \quad t\in[0,1].
\]
At $t=0$, the third and fourth points coincide.
For every $t>0$, the fourth point is distinct, while the duplicate pair at the origin is unchanged. Hence the uniqueness axiom requires
\[
D(X(t))>D(X(0)) \quad\text{for every }t>0.
\]
At the endpoints $t=0$ and $t=1$, exact GKD increases from $1.24254$ to $1.24807$.
Average pairwise distance decreases from $0.33333$ to $0.29167$, the Vendi Score decreases from $1.25047$ to $1.18768$, and RKE decreases from $0.11721$ to $0.08123$.
The determinant DPP score is zero throughout because the persistent duplicate pair makes the RBF similarity matrix singular.

\begin{table}[t]
\centering
\caption{
Numerical endpoint behavior in the controlled axiom stress tests.
Relative change is computed from the first endpoint to the second.
For DPP in the uniqueness example, both endpoint values are zero, so we report $0\%$ to indicate no change.
}
\label{tab:controlled_numeric}
\begin{tabular}{lrrr}
\hline
Measure & Initial & Final & Relative change \\
\hline
\multicolumn{4}{l}{\emph{Monotonicity: $c=1.76\rightarrow1.99$}}\\
GKD                       & 1.88847 & 1.89442 & $+0.315\%$ \\
Average pairwise distance & 1.25333 & 1.33000 & $+6.12\%$ \\
Vendi Score               & 2.29378 & 2.28758 & $-0.270\%$ \\
DPP                       & 0.37543 & 0.34676 & $-7.64\%$ \\
RKE                       & 0.67950 & 0.69101 & $+1.69\%$ \\
\hline
\multicolumn{4}{l}{\emph{Uniqueness: $t=0\rightarrow1$}}\\
GKD                       & 1.24254 & 1.24807 & $+0.445\%$ \\
Average pairwise distance & 0.33333 & 0.29167 & $-12.5\%$ \\
Vendi Score               & 1.25047 & 1.18768 & $-5.02\%$ \\
DPP                       & 0       & 0       & $0\%$ \\
RKE                       & 0.11721 & 0.08123 & $-30.7\%$ \\
\hline
\end{tabular}
\end{table}

\FloatBarrier
\subsection{ImageNet/DINOv2 Stress Tests}
\label{app:exp_imagenet}

\paragraph{ImageNet representations and batching.}

We use the $50{,}000$ images in the ImageNet-1k validation set \citep{deng2009imagenet, russakovsky2015imagenet} and embed each image using DINOv2 ViT-L/14 \citep{oquab2024dinov2}.
We convert the resulting embeddings to double precision and $\ell_2$-normalize each vector.
Each of the $1{,}000$ ImageNet classes contains $50$ validation images.
Within each class, we deterministically partition the images into five disjoint batches of size $10$.
Thus, every image appears in exactly one base configuration, giving $5{,}000$ base configurations in total.

\paragraph{Bandwidth selection.}
For each image $x_i$, let
\[
r_i := \min_{\substack{j \neq i \\ j \text{ in the same batch}}} \|x_i - x_j\|,
\]
and define 
\[
r_0 := \operatorname{median}_i r_i.
\]
For the fixed batching used in our experiments, 
\[
r_0 \approx 0.7425439401.
\]
We choose the primary bandwidth $h_0$ so that two points separated by $r_0$ have RBF similarity $1/2$.
Equivalently, 
\[
h_0 := \frac{r_0}{\sqrt{2 \log 2}} \approx 0.6306587560.
\]
We use $h_0$ for the primary analysis and $0.5 h_0$ and $2 h_0$ for robustness checks.

We exhaustively apply Euclidean-realizable perturbations to these $5{,}000$ base configurations, with each perturbation designed to isolate one axiom.
The base configurations are observed image embeddings, while the perturbed configurations are valid Euclidean configurations that need not themselves correspond to embeddings of actual image batches.
We use these controlled perturbations because the axioms concern specific changes to a configuration: monotonicity compares configurations in which pairwise distances only increase, while uniqueness compares a configuration with one obtained by introducing an exact duplicate. 
Constructing such perturbations lets us isolate the relevant axiom while retaining the geometry of real learned representations.

\paragraph{Single-edge Euclidean expansion.}
Fix a batch
\[
X=(x_1,\ldots,x_{10})
\]
and an unordered pair $\{i,j\}$.
Let $H$ be the smallest affine subspace containing the other eight points.
Let $p_i$ and $p_j$ be the orthogonal projections of $x_i$ and $x_j$ onto $H$, and write
\[
x_i=p_i+u_i, \quad x_j=p_j+u_j,
\]
where $u_i$ and $u_j$ are perpendicular to the direction space of $H$.
For every anchor point $x_k\in H$,
\[
\|x_i-x_k\|^2 = \|p_i-x_k\|^2+\|u_i\|^2,
\]
and analogously for $x_j$.
Therefore, keeping $p_i, p_j$ and the residual norms $\|u_i\|, \|u_j\|$ fixed preserves every distance from $x_i$ and $x_j$ to the other eight points.

The selected distance satisfies
\[
\|x_i-x_j\|^2 = \|p_i-p_j\|^2 + \|u_i\|^2+\|u_j\|^2 - 2\langle u_i,u_j\rangle.
\]
We can therefore increase only this distance by decreasing the inner product between $u_i$ and $u_j$ while preserving their norms.
Its largest attainable value occurs when the residual vectors are antiparallel:
\[
d_{ij,\max}^2 = \|p_i-p_j\|^2 + \bigl(\|u_i\|+\|u_j\|\bigr)^2.
\]

In every batch, the eight anchor points have affine rank $7$.
Since the embeddings lie in $\sR^{1024}$, the space perpendicular to the direction space of $H$ therefore has dimension
\[
1024-7=1017.
\]
In particular, it has dimension at least two, so whenever the selected pair has positive expansion capacity, the angle between $u_i$ and $u_j$ can be varied continuously while preserving both residual norms.
Consequently, every squared distance between $d_{ij}^2$ and $d_{ij,\max}^2$ is realizable while all other pairwise distances remain unchanged.

For $\lambda\in[0,1]$, define
\[
d_{ij}(\lambda)^2 = d_{ij}^2 + \lambda \bigl(d_{ij,\max}^2-d_{ij}^2\bigr).
\]
Let $X_\lambda$ denote any Euclidean realization of the resulting distance matrix.
Since all measures considered here depend only on pairwise distances, the choice of realization does not matter.
Thus, $\lambda=0$ gives the observed configuration and $\lambda=1$ gives the largest selected-edge expansion attainable while preserving all other pairwise distances; intermediate values interpolate in squared distance.

Every one of the
\[
5{,}000\binom{10}{2}=225{,}000
\]
candidate pairs has positive expansion capacity.
Hence no pair is filtered out of the experiment.
As an implementation check, the modified endpoint distance matrices remain Euclidean to floating-point precision:
across all reconstructed endpoint matrices, the smallest eigenvalue of the corresponding Gram matrix is approximately $-8.6 \times 10^{-14}$, and the maximum error after reconstructing pairwise distances from Euclidean coordinates is below $1.8 \times 10^{-15}$.

The parameter $\lambda$ is the fraction of the available
\emph{squared-distance} expansion, not the relative increase in the distance itself.
Define
\[
c_{ij} = \frac{d_{ij,\max}^2-d_{ij}^2}{d_{ij}^2}.
\]
Then
\[
\frac{d_{ij}(\lambda)}{d_{ij}} = \sqrt{1+\lambda c_{ij}},
\]
so the actual relative increase in the selected distance is
\[
\frac{d_{ij}(\lambda)-d_{ij}}{d_{ij}} = \sqrt{1+\lambda c_{ij}}-1.
\]
Across all $225{,}000$ edges, the median distance increases for
\[
\lambda=0.1,0.25,0.5,1
\]
are respectively
\[
4.6\%,\quad 11.1\%,\quad 21.1\%,\quad 39.1\%.
\]

\paragraph{Monotonicity reversal rates.}
For each $\lambda$, we compare the diversity of the observed configuration with that of its corresponding single-edge expansion.
A \emph{reversal} occurs when the diversity score decreases even though one pairwise distance has increased and every other pairwise distance is unchanged.

\begin{figure*}[!t]
    \centering
    \includegraphics[width=0.8\textwidth]{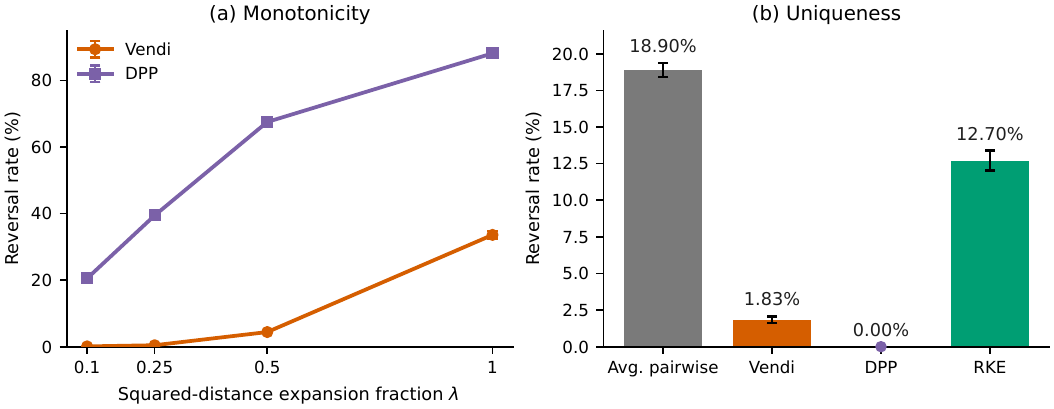}
    \caption{
    \textbf{Axiom stress tests on ImageNet/DINOv2 geometry.}
    \textbf{(a)} Reversal rates under exhaustive single-edge
    Euclidean expansions.
    \textbf{(b)} Reversal rates under all $450{,}000$ ordered duplicate replacements.
    Error bars are $95\%$ percentile bootstrap intervals from resampling ImageNet classes.
    }
    \label{fig:real_axiom_violations}
\end{figure*}

Table~\ref{tab:imagenet_monotonicity} and Figure~\ref{fig:real_axiom_violations}(a) summarize the observed reversal rates at the primary bandwidth $h_0$.
Average pairwise distance and RKE exhibit no reversals.
DPP reversals are already common at the smallest perturbation and become increasingly frequent as the expansion grows.
Vendi reversals are rare for small expansions but become frequent at the largest expansion.

\begin{table}[t]
\centering
\caption{
Observed monotonicity reversal rates at the primary bandwidth $h_0$.
Each column contains all $225{,}000$ single-edge expansions at the indicated value of $\lambda$.
}
\label{tab:imagenet_monotonicity}
\begin{tabular}{lrrrr}
\hline
Measure & $\lambda=0.1$ & $\lambda=0.25$ & $\lambda=0.5$ & $\lambda=1$ \\
\hline
Average pairwise distance & $0\%$ & $0\%$ & $0\%$ & $0\%$ \\
Vendi Score & $0.089\%$ & $0.438\%$ & $4.40\%$ & $33.61\%$ \\
DPP         & $20.53\%$ & $39.46\%$ & $67.57\%$ & $88.21\%$ \\
RKE         & $0\%$ & $0\%$ & $0\%$ & $0\%$ \\
\hline
\end{tabular}
\end{table}

The DPP reversals also become more severe as the perturbation grows.
For each reversed comparison, we measure the relative decrease
\[
\frac{
D_{\mathrm{DPP}}(X)-D_{\mathrm{DPP}}(X_\lambda)}{D_{\mathrm{DPP}}(X)}.
\]
Conditional on a reversal, the median relative decreases at $\lambda=0.1,0.25,0.5,1$ are respectively
\[
0.10\%,\quad 0.34\%,\quad 1.12\%,\quad 3.71\%.
\]
At $\lambda=1$, $65.7\%$ of all tested expansions decrease DPP by at least $1\%$, and $39.7\%$ decrease it by at least $5\%$.
Vendi's large-expansion reversals are substantially milder: at $\lambda=1$, the median decrease among reversed comparisons is $0.045\%$, and no tested comparison decreases Vendi by as much as $1\%$.

\paragraph{Uniqueness perturbation.}
For every ordered pair $i\neq j$ in each batch, we replace $x_i$ by an exact copy of $x_j$.
The observed configuration contains distinct embeddings, while the modified configuration contains one additional duplicate. The uniqueness axiom therefore requires the observed configuration to have strictly larger diversity.

There are
\[
5{,}000\cdot10\cdot9  =450{,}000
\]
ordered replacements. Figure~\ref{fig:real_axiom_violations}(b) summarizes the corresponding uniqueness reversal rates.
At the primary bandwidth $h_0$, the observed reversal rates are $18.90\%$ for average pairwise distance, $1.83\%$ for Vendi, $0\%$ for DPP, and $12.70\%$ for RKE.
There are no numerically unresolved ties at this bandwidth, so the reversal rates coincide with the observed strict-axiom failure rates.

The reversals can also be substantial.
Among reversed comparisons, the median wrong-way changes are $1.52\%$ for average pairwise distance, $1.62\%$ for Vendi, and $2.70\%$ for RKE.
Across all $450{,}000$ ordered replacements, the fractions producing a reversal of at least $1\%$ are respectively
\[
11.83\%,\quad 1.21\%,\quad 9.70\%.
\]

\paragraph{Bandwidth robustness.}
For the kernel-based measures, we repeat the experiment with the pre-specified bandwidths $0.5h_0$ and $2h_0$. 
Tables~\ref{tab:imagenet_bandwidth_monotonicity} and \ref{tab:imagenet_bandwidth_uniqueness} give the complete reversal rates.
Although the precise frequencies vary with bandwidth, the qualitative failure modes persist.

\begin{table}[t]
\centering
\caption{
Observed monotonicity reversal rates across RBF bandwidths.
Each entry is the percentage of $225{,}000$ single-edge expansions producing a wrong-direction change. Average pairwise distance and RKE have zero reversals throughout.
}
\label{tab:imagenet_bandwidth_monotonicity}
\begin{tabular}{llrrr}
\hline
Measure & $\lambda$ & $0.5h_0$ & $h_0$ & $2h_0$ \\
\hline
Vendi & $0.1$  & $0.334\%$ & $0.089\%$ & $0.004\%$ \\
      & $0.25$ & $0.877\%$ & $0.438\%$ & $0.029\%$ \\
      & $0.5$  & $2.58\%$  & $4.40\%$  & $1.09\%$  \\
      & $1$    & $6.94\%$  & $33.61\%$ & $62.24\%$ \\
\hline
DPP   & $0.1$  & $6.02\%$  & $20.53\%$ & $33.96\%$ \\
      & $0.25$ & $11.54\%$ & $39.46\%$ & $61.46\%$ \\
      & $0.5$  & $20.24\%$ & $67.57\%$ & $89.37\%$ \\
      & $1$    & $28.77\%$ & $88.21\%$ & $99.50\%$ \\
\hline
\end{tabular}
\end{table}

\begin{table}[!tb]
\centering
\caption{
Observed uniqueness reversal rates across RBF bandwidths.
Average pairwise distance is bandwidth-independent and reverses $18.90\%$ of the ordered replacements.
}
\label{tab:imagenet_bandwidth_uniqueness}
\begin{tabular}{lrrr}
\hline
Measure & $0.5h_0$ & $h_0$ & $2h_0$ \\
\hline
Vendi Score & $0.008\%$ & $1.83\%$ & $8.96\%$ \\
DPP         & $0\%$       & $0\%$      & $0\%$ \\
RKE         & $0.74\%$  & $12.70\%$ & $26.18\%$ \\
\hline
\end{tabular}
\end{table}

\paragraph{Class-level bootstrap summaries.}
The reversal rates above are exact descriptive proportions for the fixed ImageNet/DINOv2 configurations and perturbations used in the experiment.
To summarize how these rates vary with the composition of ImageNet classes, we additionally compute class-level percentile bootstrap intervals. 
We use the ImageNet class as the resampling unit: all batches and perturbations belonging to a class are kept together whenever that class is selected.
 
For each of $2{,}000$ bootstrap replicates, we sample the $1{,}000$ ImageNet classes with replacement and recompute the mean class-level reversal rate.
We report the $2.5$th and $97.5$th percentiles of the resulting bootstrap distribution.

At $\lambda=1$, the resulting $95\%$ percentile bootstrap intervals are $[87.50\%,88.86\%]$ for DPP and $[32.50\%,34.71\%]$ for Vendi.
For the uniqueness perturbation, the corresponding intervals are $[18.40\%,19.37\%]$ for average pairwise distance, $[1.61\%,2.07\%]$ for Vendi, and $[12.02\%,13.39\%]$ for RKE.

\FloatBarrier
\section{Hardness of Existing Diversity Measures on Euclidean Inputs}
\label{app:euclidean_hardness}

\citet{mironov2025measuring} construct two diversity functions, MultiDimVolume and IntegralMaxClique, that satisfy all three of their axioms but are NP-hard to compute on general dissimilarity matrices.
In this section, we show that both measures remain NP-hard when the input is restricted to Euclidean-realizable distance matrices.
Moreover, unless $\mathrm{NP} = \mathrm{RP}$, neither measure admits an FPRAS on Euclidean-realizable inputs.
The key observation is that an arbitrary graph can be encoded by a sufficiently small perturbation of the distances of a regular simplex.

\subsection{A Euclidean Graph Encoding}

\begin{lemma}[Near-equilateral Euclidean realization] \label{lem:euclidean_graph_encoding}
    Let $G = ([n], E)$ be a simple graph with $n \ge 2$, and let $0 < \eps \le 1/n$.
    Define the symmetric matrix $D = (d_{ij})$ by $d_{ii} = 0$ and, for $i \neq j$,
    \[
    d_{ij} = \begin{cases}
        2 + \eps, & \{i, j\} \in E, \\
        2, & \{i, j\} \notin E.
    \end{cases}
    \]
    Then $D$ is a Euclidean distance matrix: there exist $\vx_1, \dots, \vx_n \in \sR^{n-1}$ such that $\|\vx_i - \vx_j\| = d_{ij}$ for all $i, j$.
\end{lemma}
\begin{proof}
    Let $A$ be the adjacency matrix of $G$, let
    \[
    H := I - \frac{1}{n} \vone \vone^T,
    \]
    and define
    \[
    q := (2 + \eps)^2 - 4 = 4\eps + \eps^2.
    \]
    Consider the symmetric matrix
    \[
    B := 2H - \frac{q}{2} HAH.
    \]
    We first show that $B$ is positive semidefinite.
    For any $\vv \in \sR^n$, let $\vy := H\vv$. Since $H$ is the orthogonal projection onto $\vone^\perp$,
    \[
    \vv^T B\vv = 2\|\vy\|^2 - \frac{q}{2} \vy^T A\vy.
    \]
    Since $A$ is a $0$-$1$ adjacency matrix,
    \[
    \|A\|_1 \le n-1, \quad \|A\|_\infty \le n-1.
    \]
    Hence
    \[
    \|A\|_2 \le \sqrt{\|A\|_1 \|A\|_\infty} \le n-1.
    \]
    Therefore,
    \[
    \vy^T A\vy \le \|A\|_2 \|\vy\|^2 \le (n-1)\|\vy\|^2,
    \]
    and consequently
    \[
    \vv^T B\vv \ge \left(2 - \frac{q(n-1)}{2}\right) \|\vy\|^2.
    \]
    Because $\eps \le 1/n$,
    \[
    q(n-1) \le \left( \frac{4}{n} + \frac{1}{n^2} \right)(n-1) = 4 - \frac{3}{n} - \frac{1}{n^2} < 4.
    \]
    Thus, $B \succeq 0$.

    Since $B$ is positive semidefinite, let $r := \rank(B)$ and take a Gram factorization
    \[
        B = X^T X, \quad X \in \sR^{r \times n}.
    \]
    Let $\vx_i$ be the $i$-th column of $X$, so that
    \[
    B_{ij} = \langle \vx_i, \vx_j \rangle.
    \]
    Moreover, $H \vone = 0$ implies $B \vone = 0$, and hence
    \[
    \rank(B) \le n-1.
    \]
    Thus, we may regard $\vx_1, \dots, \vx_n$ as points in $\sR^{n-1}$.

    It remains to verify their pairwise distances. For $i \ne j$, the vector $\ve_i - \ve_j$ lies in $\vone^\perp$, so
    \[
    H(\ve_i - \ve_j) = \ve_i - \ve_j.
    \]
    Consequently,
    \[
    \|\vx_i - \vx_j\|^2 = (\ve_i - \ve_j)^T B(\ve_i - \ve_j) = 2\|\ve_i - \ve_j\|^2 - \frac{q}{2} (\ve_i - \ve_j)^T A(\ve_i - \ve_j).
    \]
    Since $G$ is simple, $A_{ii} = A_{jj} = 0$ and $A_{ij} = A_{ji}$, giving
    \[
    (\ve_i - \ve_j)^T A (\ve_i - \ve_j) = -2A_{ij}.
    \]
    Therefore,
    \[
    \|\vx_i - \vx_j\|^2 = 4 + qA_{ij}.
    \]
    If $\{i, j\} \notin E$, this equals $4$, so $\|\vx_i - \vx_j\| = 2$. If $\{i, j\} \in E$, this equals $4+q = (2+\eps)^2$, so $\|\vx_i - \vx_j\| = 2 + \eps$. Hence the resulting Euclidean configuration realizes $D$.
\end{proof}

\subsection{Exact Hardness on Euclidean Inputs}

For completeness, we recall the definition of two diversity measures of \citet{mironov2025measuring}. For a distance matrix $D = (d_{ij})$ on $n$ objects and $k \in \{2, \dots, n\}$, define
\[
M_k(D) := \max_{\substack{S \subseteq [n] \\ |S| = k}} \prod_{\substack{i, j \in S \\ i<j}} d_{ij}.
\]
The \emph{MultiDimVolume} diversity is 
\[
\operatorname{MultiDimVolume}(D) := \sum_{k=2}^n M_k(D).
\]

For $t \ge 0$, let $G_t(D)$ be the graph on $[n]$ containing an edge $\{i, j\}$ whenever $d_{ij} \ge t$, with edge weight $d_{ij}$. Among the maximum-cardinality cliques of $G_t(D)$, let $C_t$ be one with maximum total edge weight, and write
\[
w_t(D) := \sum_{\substack{i, j \in C_t \\ i<j}} d_{ij}.
\]
The \emph{IntegralMaxClique} diversity is
\[
\operatorname{IntegralMaxClique}(D) := \int_0^\infty w_t(D)\, dt.
\]

We now apply Lemma~\ref{lem:euclidean_graph_encoding} to the two axiomatic diversity measures of \citet{mironov2025measuring}.

\begin{proposition}[Hardness on Euclidean inputs] \label{prop:euclidean_hardness}
    MultiDimVolume and IntegralMaxClique are NP-hard to compute even when restricted to Euclidean-realizable distance matrices.
\end{proposition}

\begin{proof}
    We reduce from \emph{Maximum Clique}, the NP-hard problem of computing the maximum clique size $\omega(G)$ of a graph $G$ \citep{karp1972reducibility}.

    \paragraph{MultiDimVolume}

    The NP-hardness reduction of \citet{mironov2025measuring} already uses a two-distance encoding of a graph: edges have distance $2+\eps$ and nonedges have distance $2$, for a sufficiently small $\eps > 0$.
    Let $D_G$ denote this distance matrix.
    We record the quantitative details needed to ensure that the resulting instance is also Euclidean-realizable.

    For $k \in \{2, \dots, n\}$, define
    \[
    c_k := \binom{k}{2}, \quad r_k := \max_{\substack{S \subseteq [n] \\ |S| = k}} |E(G[S])|.
    \]
    For any $k$-vertex subset $S$, its product of pairwise distances is
    \[
    (2+\eps)^{|E(G[S])|} 2^{c_k - |E(G[S])|}.
    \]
    Hence the $k$-th term in MultiDimVolume is
    \[
    2^{c_k} \left(1 + \frac{\eps}{2} \right)^{r_k}.
    \]
    Writing $m := \binom{n}{2}$, choose
    \[
    \eps := 2^{-\left( 2m + \lceil \log_2 n \rceil + 3 \right)}.
    \]
    This number has $O(n^2)$ bits, thus the distance matrix $D_G$ can be constructed with polynomial bit complexity.
    It also satisfies $\eps \le 1/n$, so Lemma~\ref{lem:euclidean_graph_encoding} guarantees that the resulting distance matrix is Euclidean-realizable.

    Expanding by the binomial theorem gives
    \[
    \operatorname{MultiDimVolume}(D_G) = A_0 + \eps M + R,
    \]
    where 
    \[
    A_0 := \sum_{k=2}^n 2^{c_k}, \quad M := \sum_{k=2}^n r_k 2^{c_k-1},
    \]
    and $R \ge 0$ contains all terms of degree at least two in $\eps$.
    Since $r_k \le c_k \le m$ and $\eps/2 \le 1$,
    \[
    \begin{aligned}
        R &= \sum_{k=2}^n 2^{c_k} \sum_{j=2}^{r_k} \binom{r_k}{j} \left( \frac{\eps}{2} \right)^j \\
        &\le n \, 2^m \left( \frac{\eps}{2} \right)^2 2^m \\
        &= n \eps^2 2^{2m-2}.
    \end{aligned}
    \]
    By our choice of $\eps$, 
    \[
    0 \le \frac{R}{\eps} < \frac{1}{2}.
    \]
    Therefore the exact MultiDimVolume value determines the integer 
    \[
    M = \left\lfloor \frac{\operatorname{MultiDimVolume}(D_G) - A_0}{\eps} \right\rfloor.
    \]

    It remains to recover the values $r_k$ from $M$. Define
    \[
    w_k := 2^{c_k-1}.
    \]
    As observed in the original reduction of \citet{mironov2025measuring}, these weights satisfy
    \[
    w_k > \sum_{i=2}^{k-1} c_i w_i 
    \]
    for every $k \ge 3$. The inequality holds for $k=3$, and the inductive step follows from
    \[
    \frac{w_{k+1}}{w_k} = 2^k > \binom{k}{2} + 1.
    \]
    Since $0 \le r_k \le c_k$, we can therefore recover $r_n, r_{n-1}, \dots, r_2$ greedily from 
    \[
    M = \sum_{k=2}^n r_k w_k.
    \]
    Moreover, $M$ has $O(n^2)$ bits, so this greedy recovery requires only polynomial-time integer arithmetic.
    Finally,
    \[
    r_k = c_k
    \]
    if and only if $G$ contains a clique of size $k$. Hence
    \[
    \omega(G) = \max \left\{ k \;\middle|\; r_k = \binom{k}{2} \right\},
    \]
    so exact computation of MultiDimVolume on these Euclidean inputs is NP-hard.

    \paragraph{IntegralMaxClique.}
    For this reduction, set $\eps := 1/n$ and define $D_G = (d_{ij})$ where $d_{ii} = 0$, and for $i \ne j$,
    \[
    d_{ij} = \begin{cases}
        2 + \eps, & \{i, j\} \in E, \\
        2, & \{i, j\} \notin E.
    \end{cases}
    \]
    By Lemma~\ref{lem:euclidean_graph_encoding}, this distance matrix is Euclidean-realizable. 

    Let $S := \sum_{i<j} d_{ij}$ and let $s := \omega(G)$.
    Recall that IntegralMaxClique constructs, for each threshold $t \ge 0$, the graph containing an edge $\{i, j\}$ whenever $d_{ij} \ge t$, and lets $w_t(D_G)$ be the total edge weight of a maximum-cardinality clique, breaking ties by total weight.

    For $0 \le t \le 2$, the threshold graph is complete, so
    \[
    w_t(D_G) = S.
    \]
    For $2 < t \le 2+\eps$, the threshold graph is exactly $G$. Its maximum-cardinality clique has $s$ vertices, and every surviving edge has weight $2+\eps$, so
    \[
    w_t(D_G) = (2+\eps) \binom{s}{2}.
    \]
    For $t > 2+\eps$, the threshold graph has no edges, so
    \[
    w_t(D_G) = 0.
    \]
    Consequently, 
    \[
    \operatorname{IntegralMaxClique}(D_G) = 2S + \eps(2+\eps) \binom{s}{2}.
    \]
    Since $S$ and $\eps$ are known from the constructed instance, exact evaluation of IntegralMaxClique determines
    \[
    \binom{s}{2} = \frac{\operatorname{IntegralMaxClique}(D_G) - 2S}{\eps(2+\eps)},
    \]
    and therefore determines $s = \omega(G)$.

    Thus, IntegralMaxClique is also NP-hard to compute on Euclidean-realizable inputs.
\end{proof}

\subsection{Hardness of Randomized Approximation}
\label{sec:hardness-randomized-apx}
We next show that allowing randomized relative approximation does not make either of the two measures efficiently computable.
For a graph $G$, write $\omega(G)$ for its clique number.
We use the standard notion of a fully polynomial randomized approximation scheme (FPRAS) \citep{motwani1995randomized}: given an instance with value $F > 0$ and parameters $\eta, \delta \in (0, 1)$, the algorithm returns an estimate $\widehat F$ satisfying
\[
\Pr \! \left[ (1-\eta)F \le \widehat F \le (1+\eta) F \right] \ge 1-\delta,
\]
in time polynomial in the input size, $1/\eta$, and $\log(1/\delta)$.

\begin{theorem}[Nonexistence of FPRAS on Euclidean inputs]
\label{thm:euclidean_no_fpras}
    Unless $\mathrm{NP} = \mathrm{RP}$, neither MultiDimVolume nor IntegralMaxClique admits an FPRAS, even when restricted to Euclidean-realizable distance matrices.
\end{theorem}
\begin{proof}
    We first show that an FPRAS for either measure will allow us to recover the maximum clique size $\omega(G)$ of an arbitrary graph with constant success probability.

    \paragraph{MultiDimVolume.}
    Let $G = ([n], E)$ with $n \ge 2$, and write $\omega := \omega(G)$.
    Set $r := 100n$.
    Construct a graph $H$ by replacing each vertex $i \in [n]$ with a clique $V_i$ of $r$ vertices. 
    For distinct $i, j$, connect every vertex of $V_i$ to every vertex of $V_j$ if $\{i, j\} \in E$, and place no edges between the two clusters otherwise.
    Then
    \[
    N := |V(H)| = rn, \quad \omega(H) = r\omega.
    \]
    Expanding the maximum clique of $G$ gives a clique of size $r \omega$ in $H$.
    Conversely, the clusters intersected by any clique of $H$ correspond to a clique of $G$, so any clique of $H$ contains vertices from at most $\omega$ clusters and therefore has size at most $r\omega$.

    Define
    \[
    \rho := 1 - \frac{1}{2N+1},
    \]
    and construct a distance matrix $D_H$ by defining $d_{uu} = 0$, and for distinct $u, v$,
    \[
    d_{uv} = \begin{cases}
        1, & \{u, v\} \in E(H), \\
        \rho, & \{u, v\} \notin E(H).
    \end{cases}
    \]
    This matrix is Euclidean-realizable, since multiplying every distance by $2/\rho$ gives distance $2$ on nonedges and $\frac{2}{\rho} = 2 + \frac{1}{N}$ on edges.
    Lemma~\ref{lem:euclidean_graph_encoding} therefore applies to the scaled matrix, and scaling the realizing configuration by $\rho/2$ yields a realization of $D_H$.

    Let $M_k(D_H)$ be the $k$-th term of MultiDimVolume. 
    Since every distance is at most $1$,
    \[
    M_k(D_H) \le 1.
    \]
    Since $H$ contains a clique of size $r\omega$, $M_k(D_H) = 1$ for $2 \le k \le r\omega$. 

    We next bound the remaining terms. 
    Fix $s \ge 1$ such that $r \omega + s \le N$, and consider any set $Q \subseteq V(H)$ with
    \[
    |Q| = r\omega + s.
    \]
    For each $i \in [n]$, let
    \[
    a_i := |Q \cap V_i|, \quad x_i := \frac{a_i}{r}.
    \]
    Thus, $x_i$ is the fraction of vertices of $V_i$ contained in $Q$, and
    \[
    \sum_{i=1}^n x_i = \omega + \frac{s}{r}.
    \] 
    Let $q(Q)$ denote the number of unordered pairs in $Q$ that are nonadjacent in $H$, equivalently the number of nonedges in the induced subgraph $H[Q]$.
    Two vertices from distinct clusters $V_i$ and $V_j$ are nonadjacent in $H$ exactly when $\{i, j\} \notin E(G)$.
    Therefore,
    \[
    q(Q) = \sum_{\substack{1 \le i < j \le n \\ \{i, j\} \notin E(G)}} a_i a_j =
    r^2 \sum_{\substack{1 \le i < j \le n \\ \{i, j\} \notin E(G)}} x_i x_j.
    \]
    For any subset $R \subseteq [n]$, let $\overline e_G(R)$ denote the number of unordered pairs $\{i,j\} \subseteq R$ such that $\{i, j\} \notin E(G)$. 
    Then
    \[
    |R| - \overline e_G(R) \le \omega.
    \]
    Indeed, choose one endpoint from each missing edge inside $R$, and let $C$ be the set of chosen vertices.
    Then $|C| \le \overline e_G(R)$, and every pair of vertices in $R \setminus C$ is adjacent in $G$.
    Hence $R \setminus C$ is a clique, so
    \[
    \omega \ge |R \setminus C| \ge |R| - \overline e_G(R).
    \]
    Now form a random subset $R \subseteq [n]$ by independently including each vertex $i$ with probability $x_i$. 
    Then
    \[
    \E[|R|] = \sum_{i=1}^n x_i = \omega + \frac{s}{r},
    \]
    while
    \[
    \E[\overline e_G(R)] = \sum_{\substack{1 \le i < j \le n \\ \{i, j\} \notin E(G)}} x_i x_j.
    \]
    Taking expectations in the inequality above gives
    \[
    \omega + \frac{s}{r} - \sum_{\substack{1 \le i < j \le n \\ \{i, j\} \notin E(G)}} x_i x_j \le \omega.
    \]
    Therefore,
    \[
    \sum_{\substack{1 \le i < j \le n \\ \{i, j\} \notin E(G)}} x_i x_j \ge \frac{s}{r}.
    \]
    Recalling the expression for $q(Q)$, we conclude that $q(Q) \ge rs$.

    Every nonedge of $H$ contributes a factor $\rho$ to the pairwise distance product, whereas every edge contributes a factor $1$.
    Therefore,
    \[
    M_{r\omega + s}(D_H) \le \rho^{rs}.
    \]
    Writing 
    \[
    V_H := \operatorname{MultiDimVolume}(D_H),
    \]
    we obtain
    \[
    V_H = r\omega - 1 + T,
    \]
    where
    \[
    0 \le T \le \sum_{s=1}^\infty \rho^{rs} = \frac{\rho^r}{1-\rho^r}.
    \]
    Since 
    \[
    \rho = 1 - \frac{1}{2N+1} \le \exp \left( -\frac{1}{2N+1} \right),
    \]
    we have
    \[
    T \le \frac{1}{\exp(r/(2N+1)) - 1} \le \frac{2N+1}{r} = 2n + \frac{1}{r}.
    \]
    Suppose now that MultiDimVolume admits an FPRAS. Run it on $D_H$ with
    \[
    \eta := \frac{1}{10n}, \quad \delta := \frac{1}{4}.
    \]
    Since $M_k(D_H) \le 1$ for every $k$,
    \[
    V_H \le N-1 < N = rn.
    \]
    Thus, on the success event,
    \[
    |\widehat V_H - V_H| \le \eta V_H < \frac{r}{10}.
    \]
    Consequently,
    \[
    \begin{aligned}
        \left| \frac{\widehat V_H + 1}{r} - \omega \right| &\le \frac{|\widehat V_H - V_H|}{r} + \frac{T}{r} \\ 
        &< \frac{1}{10} + \frac{2n+1/r}{100n} < \frac{1}{2}.
    \end{aligned}
    \]
    Therefore, rounding $(\widehat V_H + 1)/r$ to the nearest integer recovers $\omega(G)$ exactly whenever the FPRAS succeeds. 

    Finally, $N = 100n^2$, the rational distance $\rho = 1 - 1/(2N+1)$ has polynomial bit complexity, and $1/\eta = 10n$. Hence the construction and the FPRAS call both have polynomial complexity in the size of $G$.

    \paragraph{IntegralMaxClique.}
    Let $G = ([n], E)$ with $n \ge 2$, and construct the Euclidean-realizable distance matrix $D_G$ from Proposition~\ref{prop:euclidean_hardness} using $\eps = 1/n$.
    Write
    \[
    s := \omega(G), \quad S := \sum_{i<j} d_{ij},
    \]
    and write
    \[
    I_G := \operatorname{IntegralMaxClique}(D_G).
    \]
    The proof of Proposition~\ref{prop:euclidean_hardness} gives
    \[
    I_G = 2S + a \binom{s}{2}, \quad a:=\eps(2+\eps).
    \]
    Since every distance is at most $2 + 1/n \le 5/2$,
    \[
    S \le \frac{5}{2} \binom{n}{2} < \frac{5}{4} n^2.
    \]
    Also,
    \[
    a = \frac{1}{n} \left(2 + \frac{1}{n} \right) \le \frac{5}{2n}, \quad \binom{s}{2} \le \binom{n}{2},
    \]
    so
    \[
    I_G < \frac{5}{2}n^2 + \frac{5}{2n} \frac{n^2}{2} < 4n^2.
    \]
    On the other hand,
    \[
    a \ge 2\eps = \frac{2}{n}.
    \]
    Suppose IntegralMaxClique has an FPRAS, and run it with 
    \[
    \eta := \frac{1}{8n^3}, \quad \delta := \frac{1}{4}.
    \]
    On the success event, define
    \[
    \widehat T := \frac{\widehat I_G - 2S}{a}.
    \]
    Since the true value is $T := \binom{s}{2}$, we have
    \[
    |\widehat T - T| = \frac{|\widehat I_G - I_G|}{a} \le \eta \frac{I_G}{a} < \eta \cdot 2n^3 = \frac{1}{4}.
    \]
    Rounding $\widehat T$ to the nearest integer therefore recovers $T = \binom{s}{2}$ exactly, and hence recovers $s = \omega(G)$.
    Since $1/\eta = 8n^3$, the entire procedure is a randomized polynomial-time procedure.

    \paragraph{From clique-number recovery to one-sided error.}
    It remains to convert either randomized clique-number procedure above into a one-sided-error algorithm for $\textsc{Clique}$. 
    Given a graph $G$ and an integer $k$, the decision problem $\textsc{Clique}$ asks whether $G$ contains a clique of size at least $k$. The problem is NP-complete \citep{karp1972reducibility}.
    The cases $k \le 1$ are trivial.

    The two recovery procedures above remain valid with any requested FPRAS failure probability $\delta'$, since their recovery guarantees hold whenever the corresponding FPRAS call succeeds.
    For each clique-number query below, set $\delta' = 1/(4n)$. 
    By the FPRAS running-time guarantee, this incurs only a polynomial dependence on $\log(4n) = O(\log n)$.

    Starting with $J := G$, process its current vertices one by one.
    For a current vertex $v$, compute $\omega(J-v)$ using the corresponding clique-number recovery procedure.
    If the result is at least $k$, delete $v$; otherwise, keep it.
    
    Condition on the event that every clique-number query returns the true clique number.
    If $G$ is a YES-instance, then $J$ continues to contain a $k$-clique throughout the process.
    Moreover, once a vertex $v$ is kept because $\omega(J-v) < k$, subsequent vertex deletions cannot make $\omega(J-v)$ increase to $k$.
    Thus, at termination, no remaining vertex can be deleted while preserving a $k$-clique.

    Thus, the final graph contains a $k$-clique.
    If it contained more than $k$ vertices, choose any $k$-clique $C$ in the remaining graph.
    Any vertex outside $C$ could be deleted while leaving $C$ intact, contradicting the termination condition.
    Therefore, the final graph has at most $k$ vertices. Since it contains a $k$-clique, it has exactly $k$ vertices and is complete.

    Finally, accept only if the remaining graph has exactly $k$ vertices and is complete.
    Hence a NO-instance is never accepted.
    For a YES-instance, the algorithm accepts whenever every clique-number query is correct.
    At most $n$ queries are made, and each has failure probability at most $1/(4n)$; by the union bound, the probability that any query fails is at most $1/4$. 
    Thus, a YES-instance is accepted with probability at least $3/4$.

    Hence $\textsc{Clique} \in \mathrm{RP}$. Since $\textsc{Clique}$ is NP-complete and $\mathrm{RP} \subseteq \mathrm{NP}$, this implies $\mathrm{NP} = \mathrm{RP}$.
\end{proof}

\end{document}

%% file: aistats/math_commands.tex
\usepackage{amsmath,amsfonts,bm}

\def\eqref#1{equation~\ref{#1}}

\def\1{\bm{1}}

\def\eps{{\epsilon}}

\def\ru{{\textnormal{u}}}

\def\ry{{\textnormal{y}}}

\def\rvu{{\mathbf{i}}}

\def\rvu{{\mathbf{u}}}
\def\rvv{{\mathbf{v}}}

\def\rvy{{\mathbf{y}}}

\def\vzero{{\bm{0}}}
\def\vone{{\bm{1}}}

\def\vc{{\bm{c}}}

\def\ve{{\bm{e}}}

\def\vm{{\bm{m}}}
\def\vn{{\bm{n}}}

\def\vr{{\bm{r}}}
\def\vs{{\bm{s}}}

\def\vu{{\bm{u}}}
\def\vv{{\bm{v}}}
\def\vw{{\bm{w}}}
\def\vx{{\bm{x}}}
\def\vy{{\bm{y}}}
\def\vz{{\bm{z}}}

\DeclareMathAlphabet{\mathsfit}{\encodingdefault}{\sfdefault}{m}{sl}
\SetMathAlphabet{\mathsfit}{bold}{\encodingdefault}{\sfdefault}{bx}{n}

\def\gB{{\mathcal{B}}}

\def\gD{{\mathcal{D}}}
\def\gE{{\mathcal{E}}}
\def\gF{{\mathcal{F}}}

\def\gM{{\mathcal{M}}}
\def\gN{{\mathcal{N}}}

\def\sC{{\mathbb{C}}}

\def\sR{{\mathbb{R}}}

\newcommand{\E}{\mathbb{E}}



%% file: aistats/macros.tex
\def\draft{0}  
\if\draft 1
\newcommand{\mycomment}[3]{{\color{#2}{[\textbf{#1: #3}]}}}

\newcommand{\Hoon}[1]{\mycomment{Younghun}{orange}{#1}}

\newcommand{\julian}[1]{{\color{blue}{\bf Julian:} #1}} 
\newcommand{\vihan}[1]{{\color{purple}{\bf Vihan:} #1}} 

\else

\newcommand{\julian}[1]{\ifhmode\unskip\fi}
\newcommand{\Hoon}[1]{\ifhmode\unskip\fi}
\newcommand{\vihan}[1]{\ifhmode\unskip\fi}

\fi